\documentclass[11pt]{article}
\pdfoutput=1
\usepackage[margin=1.0in]{geometry}
\usepackage{amsmath,amssymb,graphicx}
\usepackage{nccmath}
\usepackage{graphicx}
\graphicspath{{Figures/}{../Figures/}}

\usepackage{enumitem}
\usepackage[dvipsnames,svgnames]{xcolor}
\definecolor{mathone}{rgb}{0.3684, 0.5068, 0.709798} 
\definecolor{mathtwo}{rgb}{0.880722, 0.611041, 0.142051}
\definecolor{maththree}{rgb}{0.560181, 0.691569, 0.194885}

\usepackage[colorlinks=true,linktocpage=true,linkcolor=DarkBlue,citecolor=DarkGreen,urlcolor=DarkGreen]{hyperref}

\usepackage[T1]{fontenc}

\usepackage[utf8]{inputenc} 
\usepackage{upgreek}
\usepackage{amsmath, amsfonts, mathtools, amssymb}
\usepackage{physics}
\usepackage{slashed}
\usepackage{dsfont}
\usepackage{feynmp-auto}

\usepackage{subcaption}
\usepackage{tabularx}

\usepackage{tikz}
\usetikzlibrary{trees}
\usetikzlibrary{decorations.pathmorphing}
\usetikzlibrary{decorations.markings}
\usetikzlibrary{decorations.pathreplacing}

\usepackage{setspace}
\usepackage{authblk}
\usepackage[font=small,labelfont=bf]{caption}
\usepackage{cite}

\usepackage{mdframed}
\usepackage{pdflscape}
\usepackage[section]{placeins}
\usepackage[normalem]{ulem}

\newcommand{\beq}{\begin{equation}}
\newcommand{\eeq}{\end{equation}}
\def\bga{\begin{aligned}}
\def\eda{\end{aligned}}
\def\bgp{\begin{pmatrix}}
\def\edp{\end{pmatrix}}
\def\bgs{\begin{subequations}}
\def\eds{\end{subequations}}

\usepackage[greek,english]{babel}

\newcommand{\Mpl}{M_{\rm{pl}}}
\newcommand{\pre}{{\rm{pre}}}

\newcommand{\lnR}{{\ln R_{\rm rad}}}
\newcommand{\fnl}{{f_{\rm NL}^{\rm (loc)}}}
\newcommand{\fnlcurv}{{f_{\rm NL,curv}^{\rm (loc)}}}

\newcommand{\Sec}[1]{Sec.~\ref{#1}}

\newcommand{\App}[1]{Appendix~\ref{#1}}

\newcommand{\Fig}[1]{Fig.~\ref{#1}}
\newcommand{\Eq}[1]{Eq.~(\ref{#1})}

\newcommand{\Eqsto}[2]{Eqs.~(\ref{#1})-(\ref{#2})}

\definecolor{goodgreen}{rgb}{0,.6,0.4}

\usepackage{xspace}

\begin{document}

\title{\Large\textbf{Savior Curvatons and Large non-Gaussianity}}

\author{Jackie Lodman, Qianshu Lu, and Lisa Randall \\\vskip0.25em
\small \textit{Department of Physics, Harvard University, Cambridge, MA 02138, USA}}
\date{\today}
\maketitle

\begin{abstract}
\begin{spacing}{1.05}\noindent
\normalsize

Curvatons are light (compared to the Hubble scale during inflation) spectator fields during inflation that potentially contribute to adiabatic curvature perturbations post-inflation. They can alter CMB observables such as the spectral index $n_s$, the tensor-to-scalar ratio $r$, and the local non-Gaussianity $f_{\rm NL}^{\rm (loc)}$. We systematically explore the observable space of a curvaton with a quadratic potential. We find that when the underlying inflation model does not satisfy the $n_s$ and $r$ observational constraints but can be made viable with a significant contribution from what we call a savior curvaton, a large $f_{\rm NL}^{\rm (loc)}$ is inevitable. On the other hand, when the underlying inflation model already satisfies the $n_s$ and $r$ observational constraints, so significant curvaton contribution is forbidden, a large $f_{\rm NL}^{\rm (loc)}$ is possible in the exceptional case when the isocurvature fluctuation in the curvaton fluid is much greater than the global curvature fluctuation.

\end{spacing}
\end{abstract}

\tableofcontents

%%%%%%%%%%%%%%%%%%%%%%%%%%%%%%%%%%
%%%%%%%%%%%%%%%%%%%%%%%%%%%%%%%%%%
\section{Introduction}\label{sec:intro}
%%%%%%%%%%%%%%%%%%%%%%%%%%%%%%%%%%
%%%%%%%%%%%%%%%%%%%%%%%%%%%%%%%%%%

An inflationary epoch \cite{Guth:1980zm, Starobinsky:1980te, Linde:1981mu, Albrecht:1982wi, Linde:1983gd} resolves the horizon and flatness problems and explains the origin of primordial density perturbations that seed the temperature and matter fluctuations we see today. In single-field inflation models, the primordial fluctuations are predicted to be nearly Gaussian, consistent with the latest observations \cite{Planck:2018jri}. Future surveys will be able to probe non-Gaussianity at higher precision \cite{Alvarez:2014vva, Dore:2014cca, Abazajian:2019eic}, and detection of a primordial non-Gaussianity above the single-field inflation prediction would indicate a more complex structure during the inflationary era and provide us with insight into physics around the scale of the Hubble parameter during inflation.

One of the chief targets of a non-Gaussianity measurement is the curvaton model, which can potentially achieve a large primordial non-Gaussianity in reach of upcoming surveys \cite{Lyth:2001nq, Moroi:2001ct, Moroi:2002rd}. A curvaton, which we denote by $\sigma$, is a massive scalar field with a mass lighter than the Hubble parameter during inflation. Its equation of motion will be over-damped during inflation, causing its mean field value, $\bar{\sigma}$, to be constant during inflation, granting the curvaton a nontrivial energy density at the end of inflation (by assumption, the curvaton energy density is subdominant during inflation). After inflation, once the Hubble parameter decreases below the curvaton mass, it will start to behave like matter, and the curvaton energy density can grow relative to the radiation produced from inflaton decay. Eventually, the curvaton decays into the radiation bath, completing the reheating of the universe. 

The evolution of the curvaton fluid after inflation can potentially generate a much larger non-Gaussianity than expected from single-field inflation, in reach of detection by upcoming surveys \cite{Dore:2014cca, Abazajian:2019eic}.\footnote{There exists an alternative mechanism of generating large non-Gaussianity in curvaton models, proposed by \cite{Kumar:2019ebj}, where the curvaton is not the source of non-Gaussianity but serves as a messenger of the non-Gaussian cosmological collider signals from the inflationary epoch. We will not explore this possibility in this work.} More specifically, the non-Gaussianity generated by the curvaton is the so-called local-type non-Gaussianity, which is currently constrained by the Planck 2018 experiment  \cite{Planck:2018IX} to be $\fnl = -0.9 \pm 5.1,$ leaving much room for the possibility of a primordial $\fnl$ greater than the single-field inflation prediction of $\fnl \sim \mathcal{O}(0.01)$\cite{Acquaviva:2002ud, Maldacena:2002vr}. Henceforth in this paper, we use ``large $\fnl$'' to refer to $\abs{\fnl}>0.05$, where the non-Gaussianity is unambiguously distinguishable from the single-field inflation prediction. 

Critical to this work, the curvaton also alters the spectral index $n_s$ and tensor-to-scalar ratio $r$ from the single-field prediction of the underlying inflation model by contributing to the scalar curvature perturbation of the final radiation bath. All inflation models therefore fall in one of the following three categories.
\begin{itemize}
\item The inflation model by itself does not satisfy the $n_s$ and $r$ observational constraints, but adding a curvaton makes the inflation model viable again. We call the curvaton in this scenario a "savior curvaton."
\item The inflation model by itself already satisfies the $n_s$ and $r$ observational constraints, therefore the curvaton contribution to $n_s$\footnote{Curvaton can still contribute significantly to $r$, because adding a curvaton can only decrease $r$, and $r$ only has an upper bound \cite{BICEP:2021xfz}.} must be small. When it produce a potentially distinguishable non-Gaussianity, we call the curvaton a "stealth curvaton."
\item The inflation model does not satisfy the $n_s$ and $r$ observational constraints with or without significant contribution from a curvaton.
\end{itemize}
We do not discuss the third category of inflation models in this paper since they are, by definition, observationally ruled out.

We explicitly name the first two scenarios to emphasize the thus far underexplored interplay between the inflationary model constraints and curvaton phenomenology.\footnote{For earlier works in this area, see \cite{Lyth:2002my,Wands:2002bn,Langlois:2004nn,Sasaki:2006kq,Langlois:2008vk, Ichikawa:2008iq, Langlois:2011zz,Fonseca:2012cj, Vennin:2015vfa, Hardwick:2016whe}.} In this paper, we demonstrate that the savior and stealth curvatons generate large non-Gaussianity in qualitatively different ways. We show that savior curvatons necessarily generate a large non-Gaussianity barring fine-tuned exceptions. On the other hand, stealth curvatons exist only in the exceptional case when the isocurvature fluctuation in the curvaton sector is much greater than the global curvature fluctuation. The large ratio of fluctuations compensate for the small fractional energy density of the curvaton, producing large non-Gaussianity.

We will demonstrate the above conclusions both using the analytical formulas for the observables, $n_s$, $r$, and $\fnl$ and also using systematic numerical scans over the curvaton parameter space. We make several simplifying assumptions in this work:
\begin{itemize}
    \item The curvaton has a quadratic potential, $V(\sigma) = \frac{1}{2}m_{\sigma}^2 \sigma^2$, with $m_{\sigma} < H_{\rm end}$ such that the curvaton acquires near-scale-invariant Gaussian quantum fluctuations during inflation, where $H_{\rm end}$ is the Hubble parameter at the end of inflation.
    \item The inflaton potential has a quadratic minimum after inflation, meaning that the inflaton fluid is matter-like right after the end of inflation.\footnote{Results in this paper can still be applied to models where the inflaton fluid behave radiation-like after inflation, by using the case where the ``matter-like'' inflaton decays to radiation immediately after inflation, which we will discuss.}
    \item The curvaton decays through a dimension-5 operator suppressed by a scale $\Lambda$,
    \beq
    \Gamma_{\sigma} = \frac{m_{\sigma}^3}{\Lambda^2}.\label{eq:width}
    \eeq
    Unless specified otherwise, we assume $\Lambda = \Mpl$. We will discuss the effect of changing the scale $\Lambda$ and the functional form of $\Gamma_{\sigma}$ whenever it is relevant.
    \item The decay products of the curvaton and the inflaton form a single thermalized radiation bath.
    \item Direct interaction between the curvaton and the inflaton is negligible.
    \item The curvaton decays after the inflaton decays. We have investigated the alternative case and found that it always results in significantly suppressed curvaton contribution to all cosmological observables, because the curvaton has no opportunity to grow its energy density with respect to the inflaton sector. We will therefore not investigate this scenario further in this paper.
    \item The model parameters $m_\sigma$ and $\bar{\sigma}$ fulfill the following minimal consistency conditions: \begin{itemize}
    \item Curvaton decay happens before $H_{\rm BBN}$, defined by $T_{\rm BBN} \approx 10\; {\rm MeV}$.
    \item No second period of inflation is induced by the vacuum energy-like $\rho_{\sigma} = \frac{1}{2}m_{\sigma}^2 \sigma^2$ before the curvaton starts to oscillate like matter at $H = m_{\sigma}$. \label{minimal_consis_cond}\end{itemize}

\item We use the sudden-decay approximation, where any energy transfer process happens instantaneously when the Hubble scale equals the critical value associated with the transfer process.\footnote{See, e.g.,  \cite{Giudice:2000ex, Malik:2006pm, Sasaki:2006kq} for studies on the effect of finite-duration decays.} Therefore, the perturbative decay of the inflaton occurs when $H=\Gamma_\phi,$ where $\Gamma_\phi$ is the inflaton decay rate, and the perturbative decay of the curvton occurs at $H=\Gamma_\sigma$, where $\Gamma_\sigma$ is the curvaton decay rate. Perturbative decay of both the inflaton and curvaton completely converts their energy density into radiation.

\item We include the possibility of preheating, a non-perturbative process that partially converts the inflaton into radiation. We parameterize the partial conversion by $\alpha$, where
\beq
\rho_{\phi}^{\text{after preheating}} = \alpha \rho_{\phi}^{\text{before preheating}},\quad \rho_{r}^{\text{after preheating}} = (1-\alpha) \rho_{\phi}^{\text{before preheating}}.
\eeq
\item When included, preheating occurs right after inflation ends, i.e. the critical Hubble value is $H_{\pre} = H_{\rm end}$, where $H_{\rm end}$ is Hubble at the end of inflation. $H_{\pre}$ is usually close to $H_{\rm end}$ in known preheating models, because backreaction from the daughter particles quickly shuts down the resonant preheating process (e.g. \cite{Kofman:1997yn, Dufaux:2006ee}).\footnote{Mechanisms exist in the literature to circumvent backreaction, such as spillway preheating \cite{Fan:2021otj}, where the end of preheating is not necessarily close to the end of inflation. A prolonged preheating of efficiency $\alpha$ could be approximated as an instantaneous preheating that happens at the end of inflation with efficiency $<\alpha$, such that when the prolonged preheating is supposed to end, the ratio of inflaton to radiation energy density is the same. Modeling could be more complex if the prolonged preheating lead to changes in the universe's equation of state between the end of inflation and the end of preheating. We will not further investigate this scenario in this work.}
\label{assumptions}
\end{itemize}

The rest of the paper is organized as follows. \Sec{sec:basics} reviews results in the literature on how curvature perturbations are transferred to the radiation bath post-inflation and collects relevant equations for the computations of $\fnl$, $n_s$, and $r$. \Sec{sec:methods} describes the methods we use to evolve different fluids' energy densities from the end of inflation to the end of reheating, which we then feed into the equations described in \Sec{sec:basics}. The results from our parameter scans are presented in \Sec{sec:results}, which demonstrate the correlation of curvaton production of large $\fnl$ with the savior and stealth quality of the curvaton. We start \Sec{sec:results} by explaining why whether the inflation model already satisfies the $n_s$ and $r$ observational constraints determines when large $\fnl$ is inevitable in curvaton models. We then present the detailed results, in \Sec{sec:nsr_not_satisfied} for savior curvatons and in \Sec{sec:nsr_satisfied} for stealth curvatons. We conclude in \Sec{sec:conclusion}.

%%%%%%%%%%%%%%%%%%%%%%%%%%%%%%
%%%%%%%%%%%%%%%%%%%%%%%%%%%%%%
\section{Non-linear transfer of curvature perturbations}\label{sec:basics}
%%%%%%%%%%%%%%%%%%%%%%%%%%%%%%
%%%%%%%%%%%%%%%%%%%%%%%%%%%%%%

In this section, we review how the primordial  fluctuations in the inflaton and curvaton fields generated during inflation transfer to their decay products after inflation ends, assuming that both the inflaton and curvaton decay products thermalize to form one radiation bath. We wish to find the curvature perturbation in the radiation bath at the end of reheating, which is what we observe as the temperature fluctuation in the CMB. All equations in this section are known results, and readers familiar with the literature should skip this section.
%%%%%%%%%%%%%%%%%%%%%%%%%%%%%%
\subsection{Perturbations during inflation}
%%%%%%%%%%%%%%%%%%%%%%%%%%%%%%

Neglecting slow-roll corrections, since both the inflaton and the curvaton have a mass lighter than the Hubble parameter during inflation, their quantum fluctuations, $\delta \phi$ and $\delta \sigma$, have a power spectrum on spatially-flat hypersurfaces at horizon exit given by \cite{Byrnes:2006fr},
\beq
\mathcal{P}_{\delta \phi_*} = \left(\frac{H_*}{2\pi}\right)^2, \quad \mathcal{P}_{\delta \sigma_*} = \left(\frac{H_*}{2\pi}\right)^2,\label{eq:field_fluc}
\eeq
where $H_*$ is the Hubble parameter at the time of horizon exit, and the power spectrum $\mathcal{P}$ for any variable $x$ is defined by
\beq
\mathcal{P}_{x}\delta^3(\mathbf{k}-\mathbf{k}') \equiv \frac{4\pi k^3}{(2\pi)^3}\expval{x(\mathbf{k})x^*(\mathbf{k}')}.
\eeq

Given that direct inflaton interaction with the curvaton is negligible, the classical field trajectory during inflation is entirely in the inflaton direction \cite{Byrnes:2006fr}. Therefore, the curvature perturbation on uniform-density hypersurfaces \cite{Bardeen:1983qw} at horizon exit is sourced by the inflaton fluctuation, 
\beq
\zeta_{\phi} = -H\delta \phi/\dot{\phi}_*.
\eeq
The power spectrum of the curvature perturbation at horizon exit is therefore given by \cite{Starobinsky:1979ty, Mukhanov:1981xt, Hawking:1982cz, Starobinsky:1982ee, Guth:1982ec, Bardeen:1983qw, Mukhanov:1990me, Stewart:1993bc, Liddle:1994dx}
\beq
 \mathcal{P}_{\zeta_{\phi}} \simeq \frac{1}{2\Mpl^2 \epsilon_*}\left(\frac{H_*}{2\pi}\right)^2,
\eeq
where $\Mpl = 2.4\times 10^{18}$ GeV is the reduced Planck mass, $\epsilon \equiv -\frac{\dot{H}}{H^2}$ is the slow-roll parameter. We denote the curvature perturbation during inflation by $\zeta_{\phi}$ to indicate that this is the primordial curvature perturbation sourced by the inflaton perturbation (as opposed to curvature perturbation generated due to post-inflation dynamics). The curvaton fluctuations, on the other hand, source isocurvature perturbations during inflation. This isocurvature perturbation will be converted into curvature perturbation post-inflation when the curvaton decays, as we now discuss in detail.

%%%%%%%%%%%%%%%%%%%%%%%%%%
\subsection{After inflation until reheating}
\begin{figure}[h]
    \centering
    \includegraphics[width=0.49\textwidth]{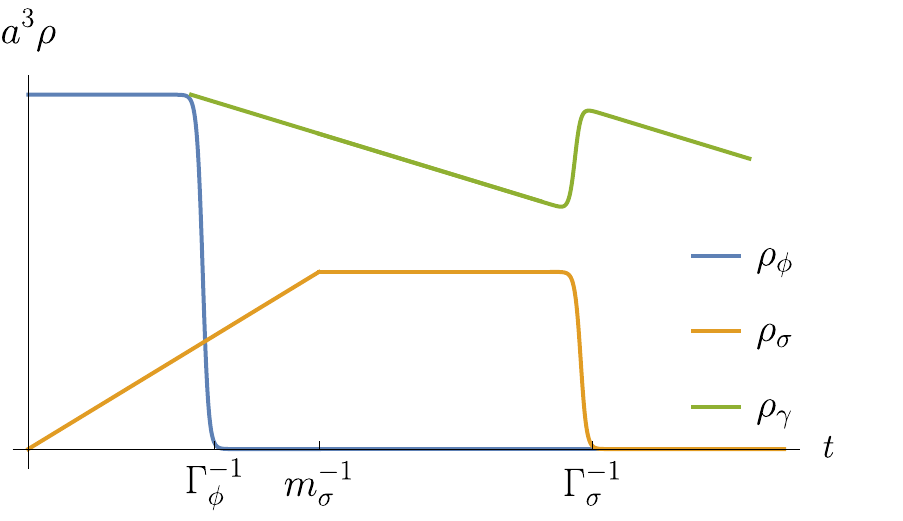}

    \caption{The schematic evolutionary history of the energy densities after inflation. Because we assume the inflaton energy density is matter-like after inflation, $a^3\rho_{\phi}$ is constant. At $H = \Gamma_{\phi}$, the inflaton decays into radiation. The radiation energy density redshifts more quickly than matter, so $a^3\rho_{\gamma}$ decreases with time. The curvaton energy density behaves as vacuum energy after inflation, so $a^3\rho_{\sigma}$ grows with time, until the curvaton starts to oscillate and behave like matter at $H = m_{\sigma}$. When $H = \Gamma_{\sigma}$, the curvaton decays into radiation. Although not explicitly drawn here, both cases $\Gamma_{\phi} > m_{\sigma} >\Gamma_{\sigma}$ and $m_{\sigma}>\Gamma_{\phi} >\Gamma_{\sigma}$ are considered in the calculation. Our calculation shows that the case of $m_{\sigma} >\Gamma_{\sigma}>\Gamma_{\phi}$ cannot generate large $\fnl$ and is not further discussed in this work.}
    \label{fig:deln0_schematic}
\end{figure}

%%%%%%%%%%%%%%%%%%%%%%%%%%
After the end of inflation, as the Hubble parameter decreases with time, the curvaton and inflaton fluids undergo different stages of evolution. By assumption, the inflaton fluid behaves as matter after inflation. The curvaton fluid, on the other hand, behaves approximately as vacuum energy when $H > m_{\sigma}$, because the Hubble friction keeps its amplitude stuck at whatever value it had during inflation. When the Hubble parameter decreases below $m_{\sigma}$, the curvaton starts to oscillate around the minimum of its quadratic potential, and the curvaton fluid also behaves as matter. Right after inflation, the inflaton will partially decay into radiation if preheating occurs, and eventually all of the (remaining) inflaton energy density decays into radiation when the inflaton perturbatively decays ($H=\Gamma_\phi$). The curvaton also decays into the same radiation bath some time after inflaton decay ($H=\Gamma_\sigma$), completing reheating. A schematic illustration of an example of the evolution of the fluids' energy densities is shown in \Fig{fig:deln0_schematic}.

We will discuss the calculation of the curvature perturbation in the case that the inflaton does not undergo preheating, as the equations without preheating are much simpler and provide important intuition that will be helpful later to understand the parameter scan results for all cases, with and without preheating. We will then briefly discuss the calculation for the case with preheating.

\textbf{Calculation of curvature perturbation without preheating:} To calculate the adiabatic curvature perturbation of the final radiation bath when reheating is completed, we use the formalism of nonlinear curvature perturbation from \cite{Lyth:2004gb}. Our notation and discussion largely follows \cite{Langlois:2011zz, Fonseca:2012cj}, which describes the application of the nonlinear curvature perturbation formalism to general multi-fluid scenarios. On superhorizon scales, the curvature perturbation of the total fluid $\zeta$ is non-linearly related to the total energy density of the universe $\rho$ by \cite{Langlois:2011zz}
\beq
\zeta = \delta N + \frac{1}{3}\int_{\bar{\rho}}^{\rho}\frac{\dd \tilde{\rho}}{(1+w)\tilde{\rho}},
\eeq
where $\delta N$ is the local perturbation in the number of expansion $e$-folds, $\bar{\rho}$ is the homogeneous energy density, and $w$ is the equation of state of the total fluid. In a multi-fluid scenario such as the curvaton model, we can analogously define individual non-linear perturbation $\zeta_i$ for each fluid species $i$,
\beq
\zeta_i = \delta N + \frac{1}{3}\int_{\bar{\rho}_i}^{\rho_i}\frac{\dd \tilde{\rho}}{(1+w_i)\tilde{\rho}}.
\eeq
Each $\zeta_i$ is conserved on super-horizon scale when there is no energy exchange between the fluids \cite{Langlois:2011zz}. However, when the universe is made of fluids with different equations of state, $\zeta$ is not conserved between the end of inflation and reheating. $\zeta$ is a weighted sum of the $\zeta_i$ \cite{Malik:2002jb}, and the relative weights change as the energy densities of the fluids evolve differently with time due to their different equations of state.

We will assume that each of our fluids has a constant equation of state. The integral is thus trivial and we obtain
\beq
\rho_i = \bar{\rho}_i \exp{3(1+w_i)(\zeta_i - \delta N)}.
\eeq

The curvature perturbation in the inflaton fluid is the known contribution at horizon exit, $\zeta_{\phi} = -H_*\delta \phi/\dot{\phi}_*$. The curvature perturbation in the curvaton fluid, on the other hand, can be calculated at the time of curvaton oscillation \cite{Fonseca:2012cj}, which we define to be $H = m_{\sigma}$. The hypersurface of $H = m_{\sigma}$ has a constant total energy density, therefore the curvature perturbation of the total fluid is simply $\zeta = \delta N$. Moreover, we assume that the inflaton is dominating the total energy density at the time of curvaton oscillation, so that $\zeta = \zeta_{\phi}$ up to when $H = m_{\sigma}$ \cite{Fonseca:2012cj}. This does not mean that the inflaton cannot convert into radiation before curvaton oscillation, either partially by preheating or completely by perturbative decay. If the inflaton-sector energy density dominates at the time of curvaton oscillation, it must also dominate before. The inflaton to radiation conversion is thus unaffected by the presence of the curvaton. The radiation created from the inflaton decay simply inherits the inflaton curvature perturbation, $\zeta_r = \zeta_{\phi}$, where $\zeta_r$ is defined as the curvature perturbation in the radiation fluid generated from inflaton decay. Therefore, the global curvature perturbation stays $\zeta = \zeta_{\phi}$ until curvaton oscillation.

The curvaton energy density at the time of oscillation can then be written as
\beq
\rho_{\sigma} = \bar{\rho}_{\sigma}e^{3(\zeta_{\sigma}-\zeta_{\phi})} = \frac{1}{2}m_{\sigma}^2\sigma_{\rm osc}^2 = \frac{1}{2}m_{\sigma}^2(\bar{\sigma}^2_{\rm osc}+\delta \sigma_{\rm osc})^2 = \left(1+\frac{\delta \sigma_{\rm osc}}{\bar{\sigma}_{\rm osc}}\right)^2\bar{\rho}_{\sigma},
\eeq
where on the first equal sign we used the definition of the nonlinear curvature perturbation in the curvaton fluid $\zeta_{\sigma}$, and in the second equal sign and beyond we simply wrote the curvaton energy density in terms of its mean amplitude at the time of oscillation $\bar{\sigma}_{\rm osc}$ and its perturbation at the time of oscillation $\delta \sigma_{\rm osc}$. For a scalar field in a quadratic potential, evolution from the classical equation of motion keeps the ratio $\delta \sigma/\bar{\sigma}$ invariant \cite{Lyth:2001nq}. Furthermore, we will assume that because of the overdamping from the Hubble friction, the mean curvaton amplitude $\bar{\sigma}$ has evolved negligibly from the time of horizon exit, $\bar{\sigma}_*$ to the time of oscillation, $\bar{\sigma}_{\rm osc}$, which we will simply denote by $\bar{\sigma}$. With the above simplifications, we obtain
\beq
e^{3(\zeta_{\sigma}-\zeta_{\phi})} = \left(1+\frac{\delta \sigma_{*}}{\bar{\sigma}}\right)^2.
\eeq
Expanding to second order, we get
\beq
3(\zeta_{\sigma}-\zeta_{\phi}) = 2\frac{\delta \sigma_*}{\bar{\sigma}} + \left(\frac{\delta\sigma_*}{\bar{\sigma}}\right)^2.\label{eq: zetasubtraction}
\eeq
Noting that $\delta\sigma_*$ generated from the quantum fluctuation of the curvaton field is a Gaussian variable, we conclude that the isocurvature perturbation in the curvaton fluid is non-Gaussian. Moreover, the non-Gaussianity of the isocurvature perturbation is the local-type, since it contains the square of a Gaussian variable. We will define 
\beq
S_{G} \equiv 2\frac{\delta \sigma_*}{\bar{\sigma}}
\eeq
to stand for the Gaussian variable. Since $S_G$ is proportional to the curvaton field fluctuation $\delta\sigma_*$ which has a power spectrum as discussed in \Eq{eq:field_fluc}, $S_G$ has a power spectrum
\beq
\mathcal{P}_{S_G} = \frac{4}{\bar{\sigma}^2}\mathcal{P}_{\delta \sigma_*} = \frac{4}{\bar{\sigma}^2}\left(\frac{H_*}{2\pi}\right)^2.
\eeq

We have already discussed how at the time of inflaton decay, the radiation produced from the inflaton simply inherits the inflaton curvature perturbation, $\zeta_{r} = \zeta_{\phi}$. Finally, when the curvaton decays at $H = \Gamma_{\sigma}$, the reheating of the universe is completed. The isocurvature perturbation in the curvaton fluid is converted to curvature perturbation, transferring the aforementioned non-Gaussianity into the curvature perturbation. The global curvature perturbation after the decay seeds the temperature fluctuation we observe today in the CMB. The global curvature perturbation is sourced by both the curvature perturbation in the inflaton fluid and the isocurvature perturbation in the curvaton fluid, given by \cite{Fonseca:2012cj}
\beq
\zeta = \zeta_{\phi} + \frac{R_{\sigma}}{3}S_{G} + \frac{R_{\sigma}}{18}\left[\frac{3}{2}-2R_{\sigma} - R_{\sigma}^2\right]S_G^2,
\eeq
where $R_{\sigma}$ is a function of the energy density ratio of the curvaton at the time of curvaton decay,
\beq
R_{\sigma} = \frac{3\Omega_{\sigma_{D\sigma}}}{4-\Omega_{\sigma_{D\sigma}}},\;\; \Omega_{\sigma_{D\sigma}} = \left.\frac{\rho_{\sigma}}{\rho_{\rm tot}}\right\rvert_{H = \Gamma_{\sigma}}. \label{eq: zeta}
\eeq

Once $\zeta$ is known in terms of the two Gaussian variables $\zeta_{\phi}$ and $S_G$, we can calculate the cosmological observables. The primordial Gaussian perturbations $\zeta_{\phi}$ and $S_G$ originate from the quantum fluctuation of two independent fields, $\phi$ and $\sigma$. Therefore, we expect zero correlation between them. The power spectrum of the global $\zeta$ is a weighted sum of the two individual power spectra \cite{Fonseca:2012cj},
\beq
\mathcal{P}_{\zeta} = \mathcal{P}_{\zeta_{\phi}}+\frac{R_{\sigma}^2}{9} \mathcal{P}_{S_G}\label{eq:Ptotal}.
\eeq
The fractional contribution of the curvaton to the total power spectrum amplitude will come into the calculations of many other observables, and it is defined to be \cite{Fonseca:2012cj}
\beq
\omega_{\sigma} \equiv \frac{\frac{R_{\sigma}^2}{9} \mathcal{P}_{S_G}}{\mathcal{P}_{\zeta}}\label{eq:omega_sig}.
\eeq
Using \Eq{eq:omega_sig}, other cosmological observables can be computed. The tensor-to-scalar ratio is \cite{Fonseca:2012cj}
\beq
r = 16\epsilon_*(1-\omega_{\sigma})\label{eq:TtoSRatio},
\eeq
where $\epsilon_* \equiv -\frac{\dot{H}}{H^2}$. The tensor-to-scalar ratio is the original single-field inflation prediction $r_{\rm single} = 16\epsilon_*$, weighted by the inflaton contribution to the curvature perturbation, $(1-\omega_{\sigma})$. If the curvature perturbation is completely sourced by the curvaton, i.e. $\omega_{\sigma} = 1$, the tensor-to-scalar ratio is zero, because the curvaton energy density is subdominant during inflation and the primordial perturbation in the curvaton field does not source tensor perturbations in the spacetime metric.

The spectral index is \cite{Fonseca:2012cj}
\beq
n_{s} - 1 =  -2 \epsilon_* + 2 \eta_{\sigma\sigma}\omega_{\sigma} + (1-\omega_{\sigma})(-4\epsilon_*+2\eta_{\phi\phi}),\label{eq:tilt}
\eeq
where $\eta_{AB} \equiv \Mpl^2 \pdv{V}{A}{B}\frac{1}{V}$, evaluated at horizon exit. The $-2\epsilon_*$ term is due to the slow-roll deviation of spacetime from scale-invariance. The $2\eta_{\sigma\sigma}\omega_{\sigma}$ and $(1-\omega_{\sigma})(-4\epsilon_*+2\eta_{\phi\phi})$ are the effect of the shape of the curvaton and inflaton potentials, each weighed by their contribution to the power spectrum. Finally, the local non-Gaussianity is computed with \cite{Langlois:2011zz}
\beq
\fnl = \left(\frac{5}{4R_{\sigma}} - \frac{5}{3} - \frac{5}{6}R_{\sigma}\right)\left(\frac{\frac{R_{\sigma}^2}{9}\mathcal{P}_{S_{G}}}{\mathcal{P}_{\zeta}}\right)^2 \equiv f_{\rm NL, curv}^{\rm(loc)}\omega_{\sigma}^2, \label{eq:fnl}
\eeq
where we defined the first factor in bracket to be $\fnlcurv$, denoting the value of $\fnl$ when $\omega_{\sigma} = 1$. 

From the above equations for $n_s$, $r$, and $\fnl$, we see that a larger fractional curvaton energy density leads to a greater $\omega_{\sigma}$, enhancing the curvaton contribution to all observables, explaining the savior curvaton generation of large non-Gaussianity. We can also see how the stealth curvaton is possible: a large $\mathcal{P}_{S_G}/\mathcal{P}_\mathcal{\zeta}$ ratio can compensate for the small $R_{\sigma}$, such that $\omega_{\sigma}\sim R_{\sigma}^2 \mathcal{P}_{S_G}/\mathcal{P}_{\zeta}$ stays small while $\fnl \sim R_{\sigma}\mathcal{P}_{S_G}/\mathcal{P}_\mathcal{\zeta}$ is large. This intuition will be helpful when we try to understand the correlations between various observables that we discover in the parameter scans. 

\begin{figure}[h]
    \centering
    \includegraphics[width=0.49\textwidth]{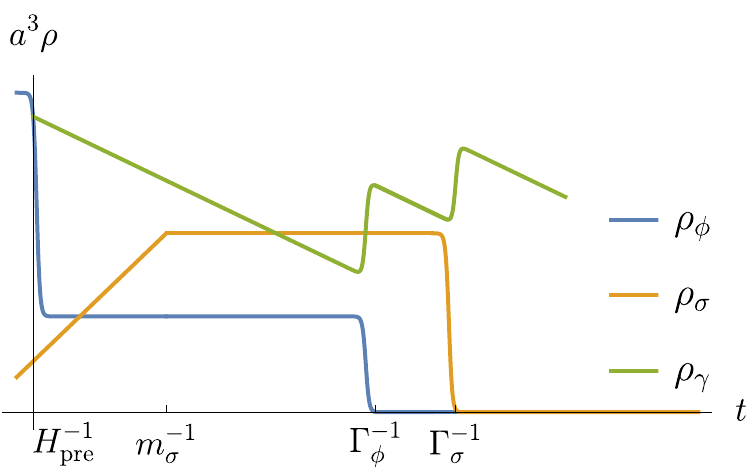}

    \caption{The schematic evolutionary history of the energy densities after inflation, assuming that the inflaton goes through a period of preheating. Right at the end of inflation, preheating converts part of inflaton into a radiation-like fluid. After preheating, the matter-like $a^3\rho_{\phi}$ is constant, while the radiation-like decay product has a decreasing $a^3\rho_{\gamma}$. At $H = \Gamma_{\phi}$, the remaining matter-like inflaton decays into radiation completely. The curvaton energy density behaves as vacuum energy after inflation, so $a^3\rho_{\sigma}$ grows with time, until the curvaton starts to oscillate and behave matter-like at $H = m_{\sigma}$. When $H = \Gamma_{\sigma}$, the curvaton decays into radiation. Although not explicitly drawn here, both cases of  $\Gamma_{\phi} > m_{\sigma} >\Gamma_{\sigma}$ and $m_{\sigma}>\Gamma_{\phi} >\Gamma_{\sigma}$ are considered in the calculation.}
    \label{fig:delnX_schematic}
\end{figure}

\textbf{Calculation of curvature perturbation with preheating:} When the inflaton goes through preheating, part of its energy density is converted to radiation immediately after inflation. The remaining matter-like inflaton then perturbatively decays later. A schematic of the energy density evolution is shown in \Fig{fig:delnX_schematic}. Preheating, since we assumed it happens right after inflation, is guaranteed to occur when the inflaton dominates the total energy density of the universe. Therefore, after preheating, the radiation-like decay product, the remaining inflaton, and the total fluid all have the primordial curvature perturbation in the inflaton, $\zeta_{\phi}$. 

Later, at the time of the perturbative decays of the inflaton and the curvaton, it is possible for any fluid to be the dominant one or for multiple fluids to have comparable energy densities, depending on model parameters, as evident from \Fig{fig:delnX_schematic}. Therefore, we need to use the general multi-fluid curvature perturbation transfer matrices calculated by \cite{Langlois:2011zz}. When a species $D$ decays into a species $A$ (for us $A$ is always the radiation bath, and $D$ is either the inflaton or the curvaton), the curvature perturbation in $A$ after the decay, up to second order in perturbation theory, is given by
\beq
\zeta_{A+} = \sum_B T_A^B \zeta_{B-} + \sum_{B, C}U_A^{BC}\zeta_{B-}\zeta_{C-}\label{eq:LL},
\eeq
where $B$ and $C$ sum over all fluid species in the universe, $+$ denote after decay and $-$ denote before decay, and $T_A^B$ and $U_A^{BC}$ are transfer matrices that depend on the equations of state and the fractional energy densities of $\{A,B\}$ and $\{A,B,C\}$. 

After both the inflaton and the curvaton have perturbatively decayed, the global curvature perturbation $\zeta$ is the curvature perturbation in the radiation fluid, $\zeta_{r}$. The curvature perturbation in the radiation fluid is calculated by using \Eq{eq:LL} twice, first with $D = \phi$ and then $D = \sigma$. The final expression of $\zeta$ is a function of $\zeta_{\phi}$, $S_{G}$, in the form \cite{Langlois:2011zz}
\beq
\zeta = \zeta_{\phi} + z_{S} S_{G} +z_{S\phi} S_{G}\zeta_{\phi}+ \frac{z_{\phi\phi}}{2}\zeta_{\phi}^2 + \frac{z_{SS}}{2} S_{G}^2,\label{eq:zeta}
\eeq
where the $z$ coefficients are functions of various ratios of energy densities at the time of the two perturbative decays, and can be found in \cite{Langlois:2011zz}.

Once $\zeta$ is known in terms of the two Gaussian variables $\zeta_{\phi}$ and $S_G$, we can calculate the cosmological observables. The fractional contribution of the curvaton to the total power spectrum amplitude is now
\beq
\omega_{\sigma} \equiv \frac{z_S^2 \mathcal{P}_{S_G}}{\mathcal{P}_{\zeta}}\label{eq:omega_sig_pre}.
\eeq
With \Eq{eq:omega_sig_pre}, the tensor-to-scalar ratio $r$ and the spectral index $n_s$ can be calculated using \Eq{eq:TtoSRatio} and \Eq{eq:tilt}. Finally, the local non-Gaussianity is computed with \cite{Langlois:2011zz}
\beq
\fnl = \frac{5}{6}\omega_{\sigma}^2\left[z_{\phi\phi}\frac{z_{\phi}^2\mathcal{P}_{\phi}^2}{z_S^2\mathcal{P}_{S_G}^2}+2z_{\phi S}\frac{z_{\phi}\mathcal{P}_{\phi}}{z_{S}\mathcal{P}_{S_G}} + z_{SS}\right].\label{eq:fnl_pre}
\eeq

%%%%%%%%%%%%%%%%%%%%%%%%%%%%%%
%%%%%%%%%%%%%%%%%%%%%%%%%%%%%%
\subsection{Determining $N_*$} \label{Sec:NstarCalc}
%%%%%%%%%%%%%%%%%%%%%%%%%%%%%%
%%%%%%%%%%%%%%%%%%%%%%%%%%%%%%

As indicated by the $*$ subscript, \Eqsto{eq:TtoSRatio}{eq:fnl} (and their counterparts for the case with preheating) are to be evaluated at the time of horizon crossing relevant to the CMB scale. We follow the Planck 2018 convention and use the pivot scale $k_0 = 0.05$ Mpc$^{-1}$, at which the amplitude of the curvature perturbation's power spectrum, $\mathcal{P}_{\zeta}$, is measured. The corresponding value of $N_*$, the number of e-folds before the end of inflation when the pivot scale left the horizon, can be computed using \cite{Liddle:2003as, Dodelson:2003vq,Martin:2010kz}
\beq
\ln R_{\rm rad} = N_* + \ln \left( \frac{k_0/a_{\rm now}}{\rho_{\gamma}^{1/4}}\right) + \frac{1}{4}\ln \rho_{\rm end} - \ln H_*.\label{eq:Nstar}
\eeq
The right-hand side of the equation computes $N_*$ assuming the universe is already radiation-dominated right at the end of inflation. When the universe is not consistently radiation dominated after the end of inflation, $\lnR$ corrects the right-hand side by accounting for those periods (between the end of inflation and reheating) where the universe was not radiation-dominated. If the universe has an average equation of state $\bar{\omega}$ when the total energy density evolved from $\rho_1$ to $\rho_2$, $\lnR$ is given by
\beq
\lnR = \frac{1-3\bar{\omega}}{12(1+\bar{\omega})}\ln\left(\frac{\rho_2}{\rho_1}\right).
\eeq
Note that $\lnR$ is additive: total $\lnR$ is the sum of all contributions from different periods of the evolution with their respective average equations of state and beginning and ending energy densities. Moreover, by definition, $\lnR$ must be zero if the average equation of state during this period is radiation-like, as evident from the $1-3\bar{\omega}$ factor. In this paper, we assumed that the curvaton does not induce a second period of inflation. Therefore, the only possible equations of state of the post-inflationary era are matter-domination or radiation-domination. Since a radiation-dominated era does not contribute to $\lnR$, $\lnR$ is a sum of contributions from all matter-dominated periods between inflation and the end of reheating.

Now, the Planck 2018 experiment has measured the scalar power spectrum amplitude to high accuracy \cite{Planck:2018jri},
\beq
\ln(10^{10} \mathcal{P}_{\zeta})=\ln\left[10^{10}\left(\mathcal{P}_{\zeta_{\phi}}+z_S^2 \mathcal{P}_{S_G}\right)\right] = 3.044 \pm 0.014.\label{eq:As_constraint}
\eeq
The number of $e$-folds obtained from \Eq{eq:Nstar} goes into the calculation of the inflaton and curvaton power spectra and the total power spectrum constraint \Eq{eq:As_constraint}. Notice that, since both \Eq{eq:Nstar} and \Eq{eq:As_constraint} depend on the scale of inflation at horizon exit, the two equations need to be solved simultaneously to obtain consistent values of $H_*$ and $N_*$. In our subsequent parameter scans, we impose the condition $\ln(10^{10} \mathcal{P}_{\zeta}) = 3.044$ on the total curvature power spectrum instead of varying over the $1$-$\sigma$ interval, given that the measured value has a high precision compared to other observables such as $n_s$, $r$, and $\fnl$.

%%%%%%%%%%%%%%%%%%%%%%%%%%%%%%
%%%%%%%%%%%%%%%%%%%%%%%%%%%%%%

%%%%%%%%%%%%%%%%%%%%%%%%%%%%%%
%%%%%%%%%%%%%%%%%%%%%%%%%%%%%%
\section{From model parameters to consistent evolution history}\label{sec:methods}
%%%%%%%%%%%%%%%%%%%%%%%%%%%%%%
%%%%%%%%%%%%%%%%%%%%%%%%%%%%%%

In the last section, we described how the cosmological observables can be computed once the evolution of the energy densities of all fluids and the equation of state of the universe are known. Here, we discuss how the energy densities and the equation of state are determined from particle physics model parameters, such as the inflaton potential, the curvaton mass, and the curvaton amplitude. We later discuss the consistency conditions imposed on the model parameters, $m_\sigma$ (curvaton mass) and $\bar{\sigma}$ (mean curvaton amplitude). 

To calculate the evolution of the energy densities and the total equation of state, we make several simplifying assumptions, as already discussed in \Sec{sec:intro}:
\begin{itemize}
    \item We use the sudden decay approximation, where any "decay" process (preheating, perturbative decays of the inflaton and curvaton) happens instantaneously when $H = H_{\pre}$ or $H = \Gamma$, where $H_{\pre}$ is the Hubble parameter when preheating ends and $\Gamma$ is the perturbative decay width. We choose $H_{\pre} = H_{\rm end}$, where $H_{\rm end}$ is the Hubble parameter at the end of inflation.

    \item The decay products of the above processes collectively form the radiation bath of the universe that exists during BBN.
    
    \item Perturbative decays of both the inflaton and curvaton completely convert their energy density into radiation, while preheating only partially converts the inflaton into radiation. We parameterize the partial conversion by $\alpha$, where
    \beq
    \rho_{\phi}^{\text{after preheating}} = \alpha \rho_{\phi}^{\text{before preheating}},\quad \rho_{r}^{\text{after preheating}} = (1-\alpha) \rho_{\phi}^{\text{before preheating}}.
    \eeq
    
    \item The equation of state of the universe is either $w = 0$ (matter-dominated, or MD) or $w = 1/3$ (radiation-dominated, or RD), according to whether matter-like or radiation-like fluids have greater energy density.
\end{itemize}

Given these simplifying assumptions, the energy densities of all fluids can be evolved using
\beq
\rho_{r} \propto a^4,\;\; \rho_{\phi} \propto a^3,\;\; \rho_{\sigma, H > m_{\sigma}} \propto a^0,\;\; \rho_{\sigma, H \leq m_{\sigma}} \propto a^3,
\eeq
\beq
H \propto a^{3/2} \text{ (MD)} \quad
  H \propto a^2 \text{ (RD)}.
\eeq
The initial conditions for the inflaton and the curvaton energy densities are set at the end of inflation, which is defined to be when $\epsilon_* = 1$, and $H_{\rm end} \equiv H(\epsilon_* = 1)$. The energy densities of the inflaton and the curvaton fluids at the end of inflation are
\beq
\rho_{\phi,{\rm end}} = \rho_{\text{total,end}} \equiv 3H_{\rm end}^2 \Mpl^2,\quad  \rho_{\sigma, {\rm end}} = \frac{1}{2}m_{\sigma}^2 \bar{\sigma}^2,
\eeq
where we have assumed that the curvaton energy density during inflation is negligible compared to the inflaton energy density.

\begin{figure}[t]
\centering
\begin{subfigure}{0.49\textwidth}
\includegraphics{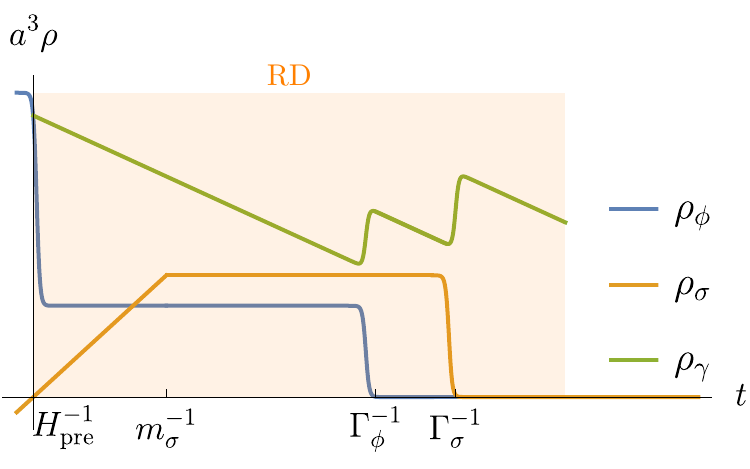}
\end{subfigure}
\begin{subfigure}{0.49\textwidth}
\includegraphics{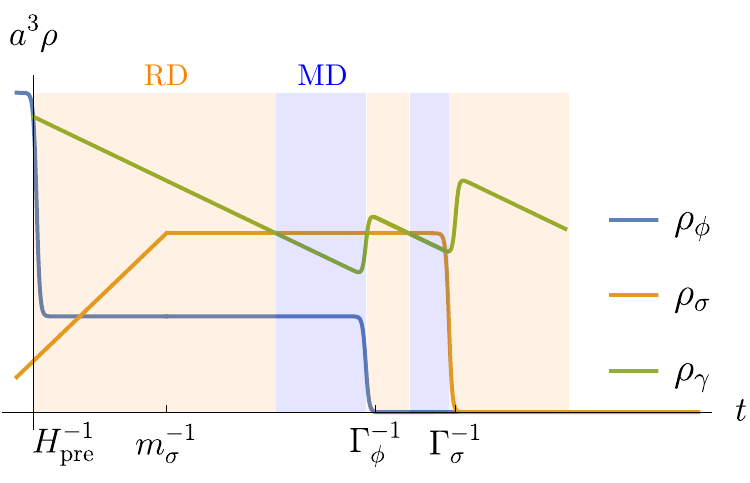}
\end{subfigure}
\caption{An example of two different equation of state transitions resulting from different values of the scale of inflation, while fixing all other model parameters. On the left side, the universe stays radiation dominated (RD) after preheating until reheating ends. On the right side, the inflaton has a lower energy density at the end of inflation, and the universe undergoes several transitions between radiation-dominated (RD) and matter-dominated (MD) phases before the end of reheating. The two evolution histories yield different dependence of $\mathcal{P}_{\zeta}$ and $N_*$ on model parameters, and we cannot know which equation of state transition corresponds to a certain model parameter choice before solving \Eq{eq:Nstar} and \Eq{eq:As_constraint} for the scale of inflation.}
\label{fig:evol_cases}
\end{figure}

As mentioned previously, we consider two possible orderings of $m_{\sigma}$, $\Gamma_{\phi}$, and $\Gamma_{\sigma}$: $\Gamma_{\phi}>m_{\sigma} > \Gamma_{\sigma}$ and $m_{\sigma} > \Gamma_{\phi} > \Gamma_{\sigma}$. The two orderings need to be treated separately, since the order of the transfer of curvature perturbations will be different. Moreover, for a particular ordering of $m_{\sigma}$, $\Gamma_{\phi}$, and $\Gamma_{\sigma}$, depending on the curvaton and inflaton parameters, the equation of state of the universe after inflation can go through several transitions from matter-dominated to radiation-dominated and vice versa. We have assumed that the universe is matter-dominated right after inflation ends. Depending on the amount of radiation injected by preheating and perturbative decay of the inflaton, the universe can become radiation dominated after each of the injection processes. In between each radiation-injecting process, given enough redshift, the universe might become matter-dominated again. Whether the universe will go through multiple such transitions of the equation of state depends on model parameters, as illustrated in \Fig{fig:evol_cases}. 

When there is preheating, how the equation of state of the universe evolves after inflation cannot be determined a priori. Different equation of state evolution histories yield different dependence of $\mathcal{P}_{\zeta}$ and $N_*$ on model parameters such as $m_{\sigma}$, $\bar{\sigma}$, $\Gamma_{\phi}$, $\alpha$. For example, assume that we have chosen certain values for $m_{\sigma}$, $\bar{\sigma}$, $\Gamma_{\phi}$, and $\alpha$. Then, the scale of inflation is not a free parameter, because it needs to satisfy the observed value of the scalar perturbation power spectrum. However, without knowing the scale of inflation, we cannot determine the transition history of the equation of state nor the forms of \Eq{eq:Nstar} and \Eq{eq:As_constraint}, because they depend on the normalization of the inflaton energy density with respect to the curvaton energy density, as shown in \Fig{fig:evol_cases}. In our parameter scans, therefore, we propagate the universe after inflation assuming all possible cases of equation of state transitions, then pick out the single internally consistent transition history after the scale of inflation and $N_*$ are determined using \Eq{eq:Nstar} and \Eq{eq:As_constraint}. When there is no preheating and the inflaton decays only perturbatively, the equation of state transitions can be determined using model parameters without solving for the scale of inflation, and there is no such complication in the calculation \cite{Vennin:2015vfa, Hardwick:2016whe}.

Having set up the complete calculation pipeline from model parameters to cosmological observables, we are ready to scan over the curvaton parameter space. We will take a UV-agnostic point of view and scan over all values of the curvaton mass and curvaton oscillation amplitude, imposing only minimal consistency conditions on the range of parameters (see \Sec{sec:intro}). The inflaton decay width is scanned over the full allowed range, $H_{\rm end} = H_{\pre} > \Gamma_{\phi} > \Gamma_{\sigma}$. 
%%%%%%%%%%%%%%%%%%%%%%%%%%%%%%
%%%%%%%%%%%%%%%%%%%%%%%%%%%%%%

\section{Measurable $\fnl$ from curvaton models}\label{sec:results}

In this section, we discuss when large $\fnl$ is inevitable in curvaton models, and how the curvaton production of large $\fnl$ is correlated with the inflation model predictions of $n_s$ and $r$. As stated in the introduction, by "large $\fnl$", we specifically mean $|\fnl| > 0.05$, such that it is unambiguously distinguishable from the single-field prediction of $\fnl \sim \mathcal{O}(0.01)$\cite{Acquaviva:2002ud, Maldacena:2002vr}. The value $0.05$ is, of course, arbitrary, but the qualitative feature of our result is unaffected by an $\mathcal{O}(1)$ change of this numerical value.

Before diving into the details of our numerical scan results, we will first discuss here the qualitative features of the correlation between the $\fnl$, $n_s$, and $r$, which can already be seen from their analytical formulas. We will discuss the simpler formulas for the case without preheating, but the general lessons will be applicable for the case with preheating as well. Recall,
\beq
r = 16 \epsilon_* (1-\omega_{\sigma}),\label{eq:TtoSRatio2}
\eeq
\beq
n_s - 1 = -2\epsilon_* + 2\eta_{\sigma\sigma}\omega_{\sigma} + (1-\omega_{\sigma})(-4\epsilon_* + 2\eta_{\phi\phi}),\label{eq:tilt2}
\eeq
\beq
\fnl = \left(\frac{5}{4R_{\sigma}} - \frac{5}{3} - \frac{5}{6}R_{\sigma}\right)\left(\frac{\frac{R_{\sigma}^2}{9}\mathcal{P}_{S_{G}}}{\mathcal{P}_{\zeta}} \right)^2\equiv f_{\rm NL, curv}^{\rm(loc)}\omega_{\sigma}^2, \label{eq:fnl2}
\eeq
where $\omega_{\sigma} = R_{\sigma}^2\mathcal{P}_{S_{G}}/9\mathcal{P}_{\zeta}$ is the fractional curvaton contribution to the total power spectrum, $R_{\sigma} = 3\Omega_{\sigma_{D\sigma}}/(4-\Omega_{\sigma_{D\sigma}})$, and $\Omega_{\sigma_{D\sigma}}$ is the fractional curvaton energy density at the time of its decay. The single-field inflation predictions for all observables are recovered in the limit $\omega_{\sigma}\rightarrow 0$.\footnote{For $\fnl$, the limit $\omega_{\sigma}\rightarrow 0$ leads to $\fnl = 0$ in the particular formula being discussed, but it should be understood that the single-field inflation prediction is $\fnl\sim \mathcal{O}(0.01)$ from \cite{Acquaviva:2002ud, Maldacena:2002vr}.}

If the inflation model predicts a smaller $n_s$\footnote{Because we have assumed that the curvaton has a quadratic potential, the curvaton contribution to $n_s$ is always positive. See for example \cite{Fujita:2014iaa, Kawasaki:2011zi} on the phenomenology of a negative, Hubble-induced mass for the curvaton.} or a smaller $n_s$ and larger $r$\footnote{From the forms of \Eq{eq:TtoSRatio2} and \Eq{eq:tilt2}, it is clear that any curvaton contribution to the universe's energy density will always affect both $n_s$ and $r$. However, since there is only a lower bound on $r$ \cite{BICEP:2021xfz}, it is fine to lower $r$ even if the value is already within the observational constraint.} on its own than the observational constraints allow, a significant curvaton contribution $\omega_{\sigma}\approx 1$ could help make the inflation model viable again. We dub curvatons in this scenario savior curvatons. Natural inflation is an example of such an inflation model. From the form of $\omega_{\sigma}$, we see that this can be achieved either with $R_{\sigma}\approx 1$, which corresponds to a significant fractional energy density in the curvaton at the time of its decay, or a small $R_{\sigma}\ll 1$ compensated by a large ratio $\mathcal{P}_{S_G}/\mathcal{P}_{\zeta}$, which corresponds to a large perturbation in the curvaton fluid relative to the global curvature perturbation.\footnote{Everything is still in perturbative control because $\mathcal{P}_{S_{G}}/\mathcal{P}_{\zeta}$ can be much greater than 1 while both $\mathcal{P}_{S_{G}}$ and $\mathcal{P}_{\zeta}$ remain much less than 1.} When either case ($R_{\sigma}\approx 1$ or $R_{\sigma} \ll 1$) is combined with $\omega_{\sigma}\approx 1$, the resulting value of $\fnl$ will be large, unless $R_{\sigma}\approx 0.6$, which results in $\fnlcurv \approx 0$. We see that a large $\fnl$ is always generated if the curvaton contribution is needed to shift the $n_s$ and $r$ values from the single-field inflation prediction into observational bounds, barring the curvaton being tuned to have a fractional energy density at the time of its decay of $R_{\sigma} \approx 0.6$. Later, our numerical scan will reveal that, depending on the inflaton energy depletion history, $R_{\sigma} \approx  0.6$ is never possible when $n_s$ and $r$ is significantly shifted from the single-field inflation prediction.

On the other hand, if the inflation model already satisfies the $n_s$ and $r$ observational bounds with its single-field prediction, $\omega_{\sigma}$ must remain much less than 1. Surprisingly, a large $\fnl$ is still possible with $\omega_{\sigma}\ll 1$. We dub curvatons that produce large $\fnl$ in this scenario stealth curvatons. To see when a large $\fnl$ is generated, it is helpful to rewrite the $\fnl$ formula as
\beq
\fnl = \frac{\frac{R_{\sigma}}{18}\left(\frac{3}{2}-2R_{\sigma}-R_{\sigma}^2\right)\mathcal{P}_{S_G}\mathcal{P}_{\zeta}}{\mathcal{P}_{\zeta}^2}.\label{eq:fnl_alt}
\eeq
From this formula, we can see that a large $\fnl$ value is generated as long as the smallness of $R_{\sigma}$ is compensated for by a large $\mathcal{P}_{S_G}/\mathcal{P}_{\zeta}$ ratio. In other words, the small fractional energy density of the curvaton is compensated for by the isocurvature perturbation in the curvaton fluid being much greater than the global curvature perturbation. Specifically, we need $\fnl \sim R_{\sigma}\mathcal{P}_{S_G}/12\mathcal{P}_{\zeta}\gtrsim \mathcal{O}(0.01)$ while $\omega_{\sigma} = R_{\sigma}^2\mathcal{P}_{S_{G}}/9\mathcal{P}_{\zeta}\ll 1$. As we will see later in our numerical scans, only a small portion of the curvaton parameter space results in a stealth curvaton (and thus a large $\fnl$). Furthermore, from the form of \Eq{eq:fnl_alt}, when $R_{\sigma}$ is small, only a large \textit{positive} $\fnl$ is possible.

\subsection{Savior curvatons}\label{sec:nsr_not_satisfied}

In this section, we will discuss the case where the inflaton potential is such that a significant curvaton contribution to $n_s$ and $r$ is necessary to satisfy the observational constraints, i.e. the curvatons are savior curvatons. We choose, as an example, the natural inflation potential \cite{Freese:1990rb, Adams:1992bn}
\beq
V(\phi) = M^4\left[1+\cos\left(\frac{\phi}{f}\right)\right]
\eeq
with $f = 5\Mpl$. The single-field inflation predictions of this model are $n_s = \{0.948, 0.952\}$ and $r = \{0.0489,0.0312\}$ for $N_* = \{50, 60\}$. Nontrivial contribution from the curvaton is necessary to bring the inflaton model inside the 1-$\sigma$ observational constraints of $n_s \in [0.9607, 0.9691]$ \cite{Planck:2018jri} and $r < 0.036$ \cite{BICEP:2021xfz}. For visual clarity, we will henceforth focus on the distribution of $\fnl$ and $n_s$ produced by the savior curvaton, because whenever the curvaton has shifted $n_s$ inside observational constraint, the $r$ observational constraint is also satisfied.

\begin{figure}[h]
    \centering     
    \includegraphics[width=0.55\textwidth]{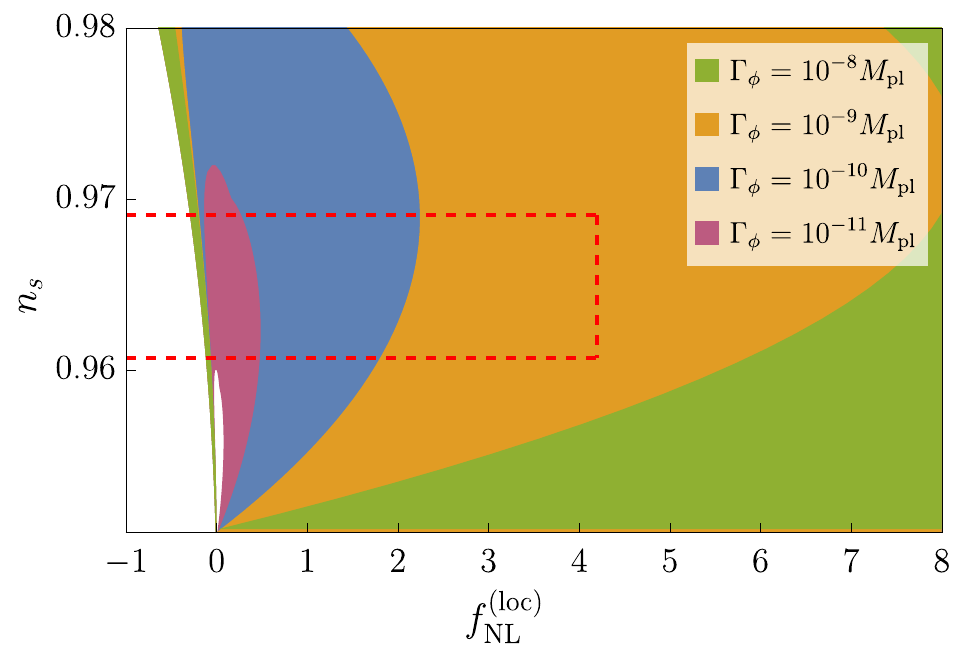}
    \caption{Distribution of $n_s$ and $\fnl$ generated by the curvaton model, assuming that the
    inflaton decays only though perturbative decay. All four colored regions overlap, with the purple containing the blue, the blue containing the orange and the orange containing the green. The dotted line is the 1-$\sigma$ allowed region in $n_s$ and $\fnl$ from the Planck 2018 constraints \cite{Planck:2018jri, Planck:2018IX}. A nontrivial upward shift in the $n_s$ value from the single-field inflation prediction of $n_s \approx 0.95$ almost guarantees a generation of $|\fnl|>0.05$, due to the hole in the $\fnl$-$n_s$ distribution around $\fnl = 0$.}
    \label{fig:deln0_ns_fnl}
\end{figure}

\Fig{fig:deln0_ns_fnl} shows the $\fnl$-$n_s$ values spanned by the curvaton model when the inflaton has a natural inflation potential with $f = 5\Mpl$ for different values of $\Gamma_{\phi}$. The single-field inflation prediction sits at the bottom-left of the plot ($\fnl \approx 0$ and $n_s \approx 0.95$). As the $n_s$ value is shifted from the single-field inflation prediction ($n_s > 0.95$), the $\fnl$-$n_s$ distribution has a hole where small $\fnl$ is prohibited. The hole sits at $\fnl = 0$ and is very close to the left edge of the $\fnl$-$n_s$ distribution, leaving only a narrow strip to the left of the hole, where $\fnl$ is negative, while to the right of the hole, the $\fnl$-$n_s$ distribution occupies a wide area where $\fnl$ is positive.

A large $\fnl$ is guaranteed for savior curvatons, as shown in \Fig{fig:deln0_ns_fnl}, if the inflaton decay width is large, and thus the hole around $\fnl = 0$ is large in the $\fnl$-$n_s$ distribution. For the particular inflation model we studied in \Fig{fig:deln0_ns_fnl}, we see that an inflaton decay width of $\Gamma_{\phi} \gtrsim 10^{-10}\Mpl$ guarantees $|\fnl|>0.05$. The size of the hole depends on the inflaton decay width because, with everything else in the model fixed, the inflaton decaying later means the energy density in the matter-like curvaton has less time to grow with respect to the radiation-like inflaton sector, suppressing the curvaton contribution to all cosmological observables. When the inflaton decay width is smaller, the hole in the $\fnl$-$n_s$ distribution is smaller, and $\fnl$ can be small even with a significantly shifted $n_s$.

\begin{figure}[h]
    \centering\includegraphics{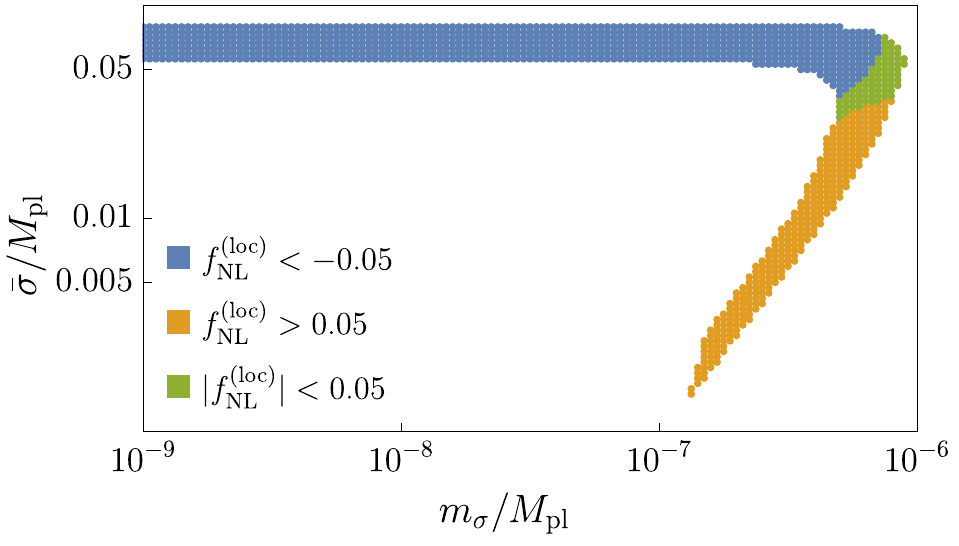}
    \caption{The distribution of $m_{\sigma}$ and $\bar{\sigma}$ that generate significant shift in $n_s$-$r$ such that the inflation model satisfies the Planck 2018 and BICEP/Keck observational constraints. The inflation model is natural inflation with $f =5\Mpl$ and the inflaton perturbative decay width is $\Gamma_{\phi} = 10^{-11}\Mpl$. The majority of the $m_{\sigma}$ and $\bar{\sigma}$ points lead to the generation of a large $\fnl$.}
    \label{fig:msig_sig_nonsr}
\end{figure}

The one exception of guaranteed large $\fnl$ is when inflaton decay width is small and $R_{\sigma} \approx 0.6$, but the latter will not be true in general. From \Fig{fig:deln0_ns_fnl}, we see that $\Gamma_{\phi} \lesssim 10^{-11}\Mpl$ is small enough for the exception to occur. However, even with a sufficiently small inflaton decay width, a small $\fnl$ does not usually result, as it corresponds to only a tiny region in the $m_\sigma$-$\bar{\sigma}$ plane. \Fig{fig:msig_sig_nonsr} shows the $m_{\sigma}$-$\bar{\sigma}$ values that generate a sufficient shift in $n_s$ and $r$ to make the inflation model observationally viable with a inflaton decay width of $\Gamma_{\phi} = 10^{-11}\Mpl$. We see that majority of the $m_{\sigma}$-$\bar{\sigma}$ values still produce a large $|\fnl|>0.05$. The small range of $m_{\sigma}$-$\bar{\sigma}$ values that generate a small $\fnl$ is exactly where the curvaton energy density at the time of decay produces $R_{\sigma} = 0.6$. The precise shape of the $m_{\sigma}$-$\bar{\sigma}$ distribution will vary with the value of $\Gamma_{\phi}$ and the functional form of the curvaton decay width. In particular, when $\Gamma_{\phi}$ decreases and the inflaton decays later, the $m_{\sigma}$-$\bar{\sigma}$ distribution will shift to smaller values of $m_{\sigma}$ so that the curvaton also decays later. Similarly, if the scale that suppresses the curvaton decay width, $\Lambda$, decreases, the $m_{\sigma}$-$\bar{\sigma}$ distribution will also shift to smaller $m_{\sigma}$. But regardless of the shifting of the $m_{\sigma}$-$\bar{\sigma}$ distribution, the fact that the majority of the $m_{\sigma}$-$\bar{\sigma}$ values lead to large $\fnl$ remains true.

The boundary of the hole in $\fnl$-$n_s$ distribution is the contour of maximum curvaton mass $m_{\sigma} = H_{\rm end}$ and varying curvaton amplitude, $\bar{\sigma}$. The fact that varying $\bar{\sigma}$ traces out a loop in the $\fnl$-$n_s$ plane can be understood as the following: when the curvaton amplitude is small, the curvaton fractional energy density at the time of decay is negligible, leading to negligible curvaton contribution to all observables, recovering the single-field inflation prediction of $\fnl \approx 0$ and $n_s \approx 0.95$. As $\bar{\sigma}$ increases, the curvaton fractional energy density at the time of decay increases, increasing curvaton contributions to observables. Thus, $\fnl$ and $n_s$ both increase. As $R_{\sigma}$ approaches and crosses $0.6$, $\fnl$ decreases and crosses 0 to become negative, while $n_s$ continues to increase. At some point, $\bar{\sigma}$ becomes so large that $R_{\sigma}\approx 1$. From there, as $\bar{\sigma}$ continues to increase, $R_{\sigma}$ can no longer increase (since its maximum value is 1), and the curvaton contribution to observables starts to decrease because $\mathcal{P}_{S_G}$, the isocurvature perturbation in the curvaton fluid, decreases as $\bar{\sigma}$ increases (specifically $\mathcal{P}_{S_G}\sim 1/\bar{\sigma}^2$). Therefore, $|\fnl|$ and $n_s$ decrease, completing the loop-like trajectory in the $\fnl$-$n_s$ back to the single-field inflation prediction.

\begin{figure}[h!]
    \centering
    \begin{subfigure}{0.55\textwidth}
    \includegraphics[width=\textwidth]{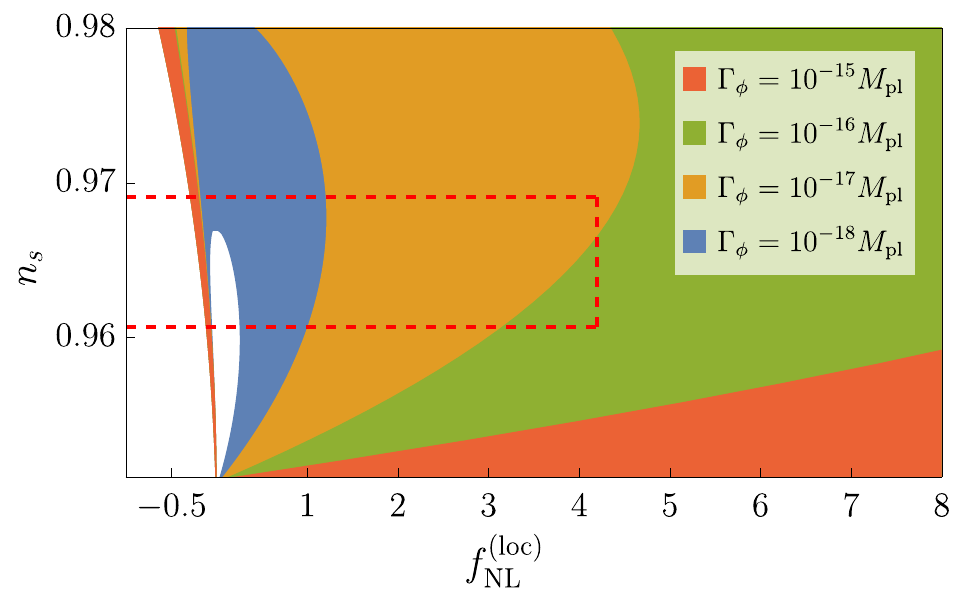}
    \end{subfigure}
    \caption{Distribution of $n_s$ and $\fnl$ in the curvaton model, assuming that the inflaton sector undergoes preheating with $\alpha = 10^{-4}$ and a perturbative decay later at $H = \Gamma_{\phi}$. All colored regions are overlapping, with the blue containing the orange, the orange containing the green, and the green containing the red. The dotted line is the 1-$\sigma$ constraint on $n_s$ and $\fnl$ from Planck 2018 \cite{Planck:2018jri, Planck:2018IX}.}
    \label{fig:deln4_ns_fnl}
\end{figure}

When we include the possibility of preheating with sufficiently small $\alpha$ at the end of inflation, the small $\fnl$ exception no longer occurs. The inflaton needs to have a sufficiently large coupling with its daughter particles to trigger the non-perturbative process of preheating, which demands a large inflaton decay width and thus a large hole in the $\fnl$-$n_s$ distribution. \Fig{fig:deln4_ns_fnl} shows the distribution of $\fnl$-$n_s$ generated in the curvaton model assuming that the inflaton undergoes a period of preheating at the end of inflation with $\alpha = 10^{-4}$. Preheating with $\alpha = 10^{-4}$ can be achieved, for example, in the spillway preheating model \cite{Fan:2021otj}. Translating the required magnitude of coupling in \cite{Fan:2021otj} to our inflation model, it corresponds to a minimal value of $\Gamma_{\phi}$ of $\Gamma_{\phi} \gtrsim 10^{-14}\Mpl$, which guarantees $\fnl$ to be large when $n_s$ is sufficiently shifted to meet the observational constraint.

Note that compared to the case without preheating, the $\fnl$-$n_s$ distribution of the same hole size now corresponds to a much smaller $\Gamma_{\phi}$ value when there is preheating. This is simply because, with everything else in the model fixed, adding preheating forces some of inflaton to be converted to radiation at the end of inflation, so the remaining matter-like inflaton must decay later to achieve the same fractional curvaton energy density at the time of curvaton decay, and thus the same $\fnl$-$n_s$ distribution shape.

\FloatBarrier

%%%%%%%%%%%%%%%%%%%%%%%%%%%%%%%%%%
%%%%%%%%%%%%%%%%%%%%%%%%%%%%%%%%%%
\subsection{Stealth curvatons}\label{sec:nsr_satisfied}

We now turn to stealth curvatons, which produce distinguishable $\fnl$ in inflation models that already satisfy $n_s$ and $r$ constraints. An example of an inflation model that satisfies the $n_s$ and $r$ observational constraints without a curvaton contribution is the $\alpha$-attractor T-model \cite{Ferrara:2013rsa, Kallosh:2013hoa, Kallosh:2013yoa},
\beq
V(\phi) = M^4 \tanh^{2n}{\frac{\phi}{\alpha\sqrt{6}}},
\eeq
with $\alpha = 0.1$ and $n = 1$ . Stealth curvatons occur when the isocurvature fluctuation in the curvaton fluid is much greater than the global curvature perturbation, generating a large positive $\fnl$. The large ratio of perturbations counter the small fractional energy density in the curvaton to generate large non-Gaussianity. In this section we clarify the conditions on the stealth curvaton parameters for generation of large $\fnl$.

\begin{figure}[h]
    \centering
    \includegraphics[width=0.6\textwidth]{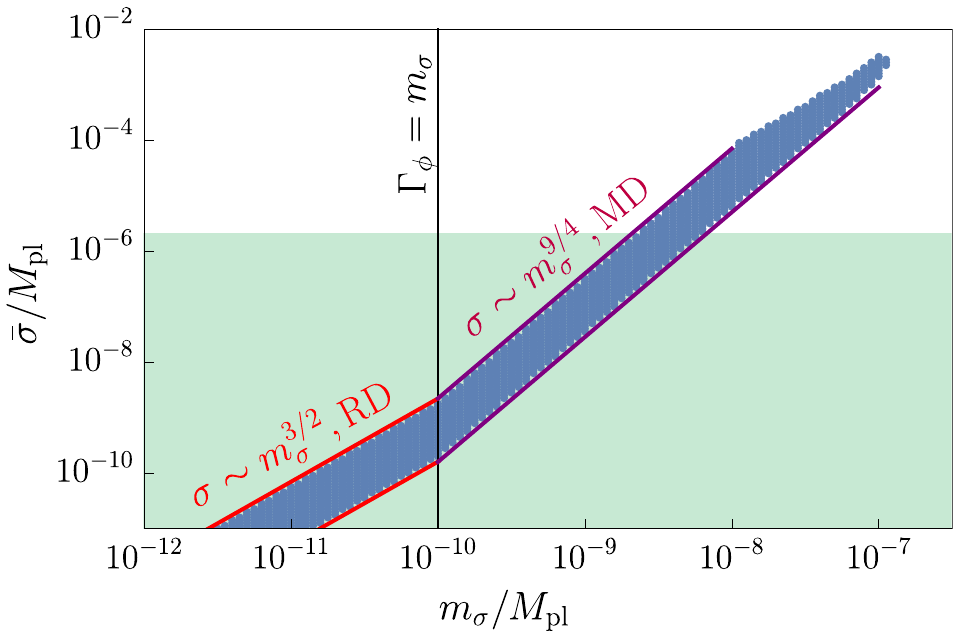}
    \caption{The distribution of $m_{\sigma}$ and $\bar{\sigma}$ that generates $0.05 < \fnl < 4.2$. All points satisfy the 1-$\sigma$ observational constraints of $n_s$ from Planck 2018 \cite{Planck:2018jri} and $r$ from BICEP/Keck 2021 \cite{BICEP:2021xfz}. The inflation model is the $\alpha$-attractor T-model with $\alpha = 0.1$ and $n = 1$, and the inflaton decay width is chosen to be $\Gamma_{\phi} = 10^{-10}\Mpl$. The distribution that corresponds to $0.05 < \fnl < 4.2$ follows $\bar{\sigma}\sim m_{\sigma}^{3/2}$ when $m_{\sigma} < \Gamma_{\phi}$ and $\bar{\sigma}\sim m_{\sigma}^{9/4}$ when $m_{\sigma} > \Gamma_{\phi}$, corresponding to curvaton beginning oscillation during radiation domination (RD) and matter domination (MD) of the universe respectively. At large $\bar{\sigma}$ and $m_{\sigma}$, the curvaton fractional energy density at the time of its decay is large, and the observational constraint on $n_s$ cuts off the distribution. The green shaded area denote regions where the curvaton fluctuation during inflation is greater than its mean amplitude, $\mathcal{P}_{\delta \sigma_*} \gtrsim \bar{\sigma}^2$. }
    \label{fig:param_gravDecay}
\end{figure}

\Fig{fig:param_gravDecay} presents the distribution of $m_{\sigma}$ and $\bar{\sigma}$ that produces a $\fnl$ between $[0.05, 4.2]$ (recall that $\fnl$ cannot be large and negative for small $R_\sigma$, as shown in \Eq{eq:fnl_alt}) and also satisfies the $1$-$\sigma$ constraint on $n_s$ from Planck 2018 \cite{Planck:2018jri} and $r$ from BICEP/Keck 2021 \cite{BICEP:2021xfz}. The upper bound of $\fnl < 4.2$ comes from the 1-$\sigma$ bound from Planck 2018 \cite{Planck:2018IX}, while the lower bound of $\fnl > 0.05$ is again chosen to ensure that the $\fnl$ is unambiguously distinguishable from single-field inflation prediction. We choose an inflaton decay width of $\Gamma_{\phi} = 10^{-10}\Mpl$ as an example. Only a narrow strip of parameter space in the $m_{\sigma}$-$\bar{\sigma}$ plane generates a large $\fnl$. In particular, all values of $m_{\sigma}$-$\bar{\sigma}$ below the blue strip in \Fig{fig:param_gravDecay} are consistent with observational constraints but only produce $|\fnl| < 0.05$.

For a wide range of $m_{\sigma}$ and $\bar{\sigma}$, the $m_{\sigma}$-$\bar{\sigma}$ distribution that produces a large $\fnl$ follows a scaling relation, with different power for $m_{\sigma} < \Gamma_{\phi}$ and $m_{\sigma}>\Gamma_{\phi}$. The difference in the power of the scaling relation comes from the whether the curvaton oscillation (defined as $H = m_{\sigma}$) happens during matter domination (MD) or radiation domination (RD): 
\begin{itemize}
    \item $m_{\sigma}<\Gamma_{\phi}$: curvaton oscillates after inflaton decay, during RD, $\bar{\sigma} \sim m_{\sigma}^{3/2}$
    \item $m_{\sigma}>\Gamma_{\phi}$: curvaton oscillates before inflaton decay, during MD, $\bar{\sigma} \sim m_{\sigma}^{9/4}$. 
\end{itemize} We will see that this correspondence with the equation of state of the universe at the time of curvaton oscillation with the scaling behavior of the $m_{\sigma}$-$\bar{\sigma}$ distribution applies to a wide range of scenarios. For more detailed derivation and discussion of the scaling relation, including an analytical derivation in certain limits, see \App{app:scaling}.

When $\bar{\sigma}$ becomes smaller than the Hubble parameter during inflation, which sets the typical fluctuation in the curvaton field, the curvature perturbation of the curvaton fluid becomes non-perturbative. This region is denoted by the green-shaded area in \Fig{fig:param_gravDecay} ($H_\text{inf}\propto 10^{-6}\Mpl$ in this model). In this region, the fluctuation in the curvaton energy density is larger than the mean curvaton energy density,
\beq
\frac{\delta \rho_{\sigma}}{\bar{\rho}_{\sigma}}  = \frac{\rho_{\sigma}-\bar{\rho}_{\sigma}}{\bar{\rho}_{\sigma}}= \frac{\frac{1}{2}(2 \bar{\sigma} \delta \sigma  + \delta\sigma ^2) m_{\sigma}^2}{\frac{1}{2}\bar{\sigma}^2 m_{\sigma}^2} = \frac{\delta \sigma}{\bar{\sigma}} + \frac{\delta\sigma^2}{\bar{\sigma}^2} > 1.
\eeq
This does not necessarily mean that the curvaton model in this regime is ruled out observationally, since the curvaton contribution to the total curvature perturbation could be suppressed, i.e. $\omega_{\sigma}\ll 1$. However, it does imply that our method of calculation is not valid, since the mean energy density of the curvaton fluid, $\bar{\rho}_{\sigma}$, which we evolve and use to calculate $R_{\sigma}$, is no longer a good indicator of the typical energy density of the curvaton fluid. However, we will not cut our numerical scan results at $\bar{\sigma} < H_*$, since the scaling behavior of the $m_{\sigma}$-$\bar{\sigma}$ distribution we find from our numerical analysis is inflation model-independent (as derived in \App{app:scaling}), and a lower-scale inflation model would allow the $m_{\sigma}$-$\bar{\sigma}$ distribution to extend to smaller $\bar{\sigma}$ values without breaking perturbativity.

\begin{figure}[h]
    \centering
    \begin{subfigure}{0.49\textwidth}
    \includegraphics[width=\textwidth]{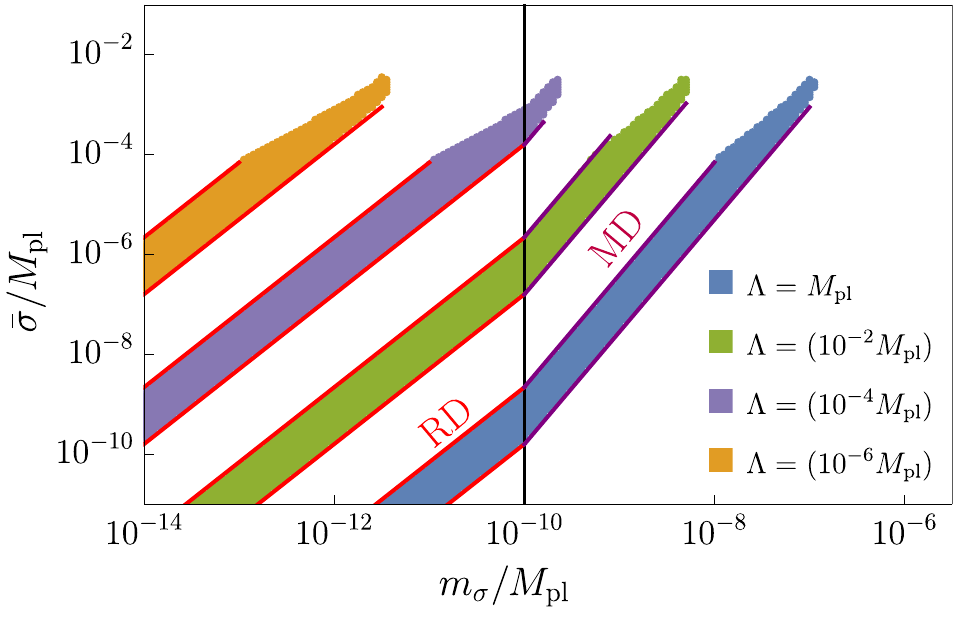}\phantomcaption\label{fig:param_varyLambdaA}
    \end{subfigure}
    \begin{subfigure}{0.49\textwidth}
    \includegraphics[width=\textwidth]{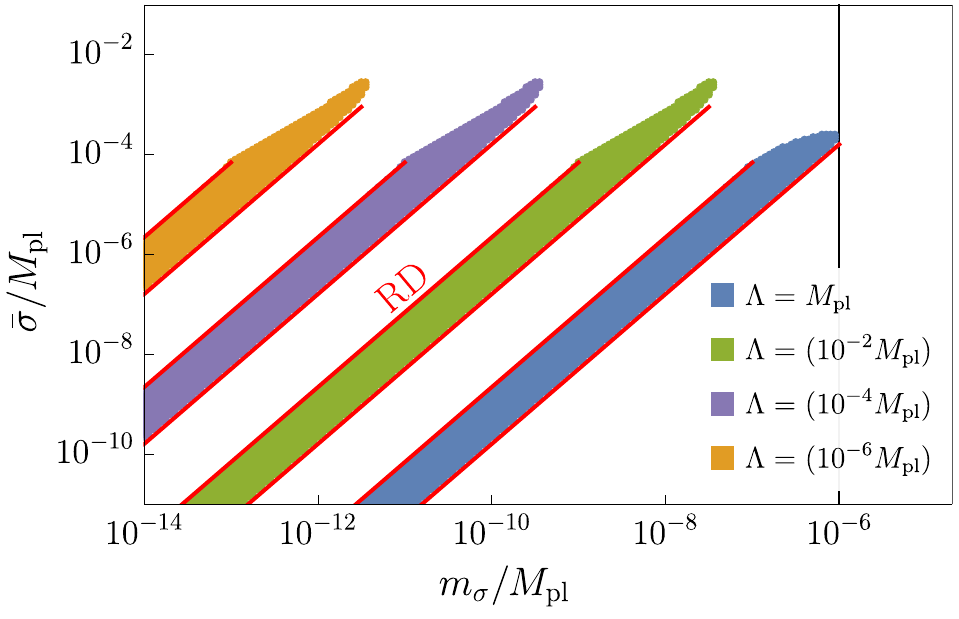}\phantomcaption\label{fig:param_varyLambdaB}
    \end{subfigure}
    \caption{The distribution of $m_{\sigma}$ and $\bar{\sigma}$ that generates $0.05 < \fnl < 4.2$ and satisfies the 1-$\sigma$ observational constraint of $n_s$ and $r$ for models with different values of $\Lambda$ that mediates curvaton decay. The inflaton decay width is set to $\Gamma_{\phi} = 10^{-10}\Mpl$ on the left and $10^{-6}\Mpl$ on the right, corresponding to different models of inflaton interaction with its daughter fields. As $\Lambda$ lowers, the distribution shifts to smaller $m_{\sigma}$ to compensate for the faster decay due to smaller $\Lambda$. For each value of $\Lambda$, the maximum possible $m_{\sigma}$ is reached when $\Gamma_{\phi} = 10^{-6}\Mpl$. This maximum value of $m_{\sigma}$ decreases with $\Lambda$ as $m_{\sigma,{\rm max}}\sim \Lambda$.}
    \label{fig:param_varyLambda}
\end{figure}

\Fig{fig:param_varyLambda} shows the $m_{\sigma}$-$\bar{\sigma}$ distribution for $\Lambda = \Mpl$, $10^{-2}\Mpl$, $10^{-4}\Mpl$, and $10^{-6}\Mpl$, where recall $\Lambda$ is the scale that suppresses the curvaton decay width in \Eq{eq:width}, and different $\Lambda$ values correspond to different models of curvaton interaction with its daughter fields. We see that models with a lower scale $\Lambda$ have $m_{\sigma}$-$\bar{\sigma}$ distributions that are more shifted to the left, to smaller $m_{\sigma}$, such that the decay width of the curvaton $\Gamma_{\sigma} = m_{\sigma}^3/\Lambda^2$ remains at similar magnitude. On the other hand, models with different inflaton decay widths have different $m_{\sigma}$-$\bar{\sigma}$ distributions that switch from the RD scaling relation to MD scaling relation at different values of $m_{\sigma}$: in \Fig{fig:param_varyLambdaA} the inflaton decay width is $\Gamma_{\phi}  = 10^{-10}\Mpl$, and the ``bend'' in the distribution occurs at $m_{\sigma} = 10^{-10}\Mpl$ for all values of $\Lambda$. In \Fig{fig:param_varyLambdaB} the inflaton decay width is $\Gamma_{\phi} = 10^{-6}\Mpl$, which is greater than the largest value of $m_{\sigma}$ allowed for all $\Lambda$, so the $m_{\sigma}$-$\bar{\sigma}$ distributions follow $\bar{\sigma} \sim m_{\sigma}^{3/2}$ throughout.

From \Fig{fig:param_varyLambdaB}, we see that because models with lower values of $\Lambda$ have $m_{\sigma}$-$\bar{\sigma}$ distributions that are more shifted to smaller $m_{\sigma}$, the maximum possible value of $m_{\sigma}$ that generates large $\fnl$ (attained when $\Gamma_{\phi}$ is the largest) is smaller for models with smaller $\Lambda$, with $m_{\sigma,{\rm max}}\sim \Lambda$. The general lesson applies for any model where the curvaton decay width is of the form $\Gamma_{\sigma} = m_{\sigma}^{2N+1}/\Lambda^{2N}$, which corresponds to the curvaton interaction with its decay products being mediated by a dimension-$(N+4)$ operator, and any inflation model. Changing $N$ changes the power in the scaling behavior, but the maximum possible value of $m_{\sigma}$ is always reached when $\Gamma_{\phi}$ is the largest, and this $m_{\sigma,{\rm max}}$ value decreases with for models with lower values of $\Lambda$, with $m_{\sigma,{\rm max}}\sim \Lambda$, i.e. constant $\Gamma_{\sigma}/m_{\sigma}$.

When preheating happens with $\alpha<0.5$, the universe will stay RD if $\Gamma_{\phi} > H_{\rm eq}$, where $H_{\rm eq}$ is when the radiation produced from preheating has equal energy density with the matter-like inflaton left after preheating,
\beq
\frac{H_{\rm eq}}{H_{\rm end}} = \left(\frac{a_{\rm end}}{a_{\rm eq}}\right)^2 = \left(\frac{\rho_{\phi}}{\rho_{\rm radiation}}\right)^2 = \left(\frac{\alpha}{1-\alpha}\right)^2.
\eeq 
For example, for $\alpha = 10^{-4}$ and $H_{\rm end}\approx 10^{-6}\Mpl$, the equation of state of the universe will be RD as long as $\Gamma_{\phi} > 10^{-14}\Mpl$, and the curvaton parameter space that generate large $\fnl$ will always scale like $\bar{\sigma}\propto m_\sigma^{3/2}$.

\section{Conclusion}\label{sec:conclusion}

The early universe presents a unique opportunity to probe physics beyond the reach of terrestrial experiments. In particular, CMB observables provide valuable information about the inflationary era at an energy scale that is potentially as high as $\mathcal{O}(10^{13})$ GeV. The local non-Gaussianity $\fnl$ is especially interesting because the single-field inflation prediction for this quantity is slow-roll suppressed, making for a clean handle on non-minimal scenarios during inflation \cite{Acquaviva:2002ud, Maldacena:2002vr}. With upcoming experiments projecting to improve sensitivity to $\fnl$ by an order of magnitude \cite{Abazajian:2019eic, Dore:2014cca}, it is increasingly crucial to understand precisely when a large non-Gaussianity is generated in different models.

In this paper, we investigated curvaton models' abilities to generate a large local non-Gaussianity and clarified when a non-Gaussianity unambiguously greater than the single-field inflation prediction is inevitable these models. We systematically scanned over the physically consistent curvaton parameter space and investigated the correlation between various cosmological observables.

We showed that if significant curvaton contribution is necessary to shift $n_s$ (or $n_s$ and $r$) from the single-field inflation prediction to meet observational bounds, the curvaton model almost always produces a large $\fnl$. We call curvatons in these models "savior curvatons." When the inflaton decay width is sufficiently small, small $\fnl$ is still possible, but it only corresponds to the small region of curvaton parameter space where the fractional curvaton energy density at the time of curvaton decay is tuned to result in $R_{\sigma} = 0.6$. This exception does not occur if the inflation decay width is large or if there is preheating in the inflaton sector with high energy transfer efficiency.

When the inflation model already satisfies the $n_s$ and $r$ observational constraints, we showed that the curvaton generically produces a small $\fnl$ that is indistinguishable from the single-field inflation prediction. Generating a large $\fnl$ requires the curvaton amplitude and mass to sit in a narrow strip of parameter space that follows a tight scaling relation, and we dubbed the curvatons that produce large $\fnl$ in these models "stealth curvatons". The power of the scaling relation indicates the equation of state of the universe when the curvaton starts to oscillate. For curvaton with a decay width in the form of $\Gamma_{\sigma} = m_{\sigma}^3/\Lambda^2$, the curvaton oscillation amplitude scaled as $\bar{\sigma}\sim m_{\sigma}^{3/2}$ for radiation domination and as $\bar{\sigma}\sim m_{\sigma}^{9/4}$ for matter domination. Different models with different values of $\Lambda$ or different values of the inflaton decay width $\Gamma_{\phi}$ will have the narrow strip of parameter space to generate large $\fnl$ at different locations in the $m_{\sigma}$-$\bar{\sigma}$ plane.

As mentioned in the introduction, we have made several simplifying assumptions in this work to isolate the effect of the curvaton on adiabatic curvature perturbations and also to stay UV-agnostic in our parameter scan. Once the curvaton is embedded in a UV-completed theory of inflation (for a review, see \cite{Mazumdar:2010sa}), the ranges of well-motivated curvaton masses and curvaton oscillation amplitudes are potentially much more limited, and details of the curvaton decay will be clearer. Moreover, if some of the curvaton decay products are not thermalized with the radiation bath of the universe, there are potentially additional constraints from the curvaton contribution to isocurvature perturbation and dark sector energy density. A study of a more UV-complete theory is necessary to understand the full scope of curvaton phenomenology, in particular, the three-way interplay between the correlation in visible sector observables we have found in this paper, restriction of parameter space from UV considerations, and the phenomenology of curvaton decay into the supersymmetric/dark sector.

\subsection*{Acknowledgments}

JL and QL would like to thank Matt Reece for helpful discussions and for feedback on previous drafts of this work. QL is grateful to Rashmish Mishra for useful discussions. JL would like to thank Daniel Green for his comments on future observational prospects of cosmological observables mentioned in this work. The work of QL is supported by the DOE grant DE-SC0013607, the Alfred P.~Sloan Foundation Grant No.~G-2019-12504, and the NASA Grant 80NSSC20K0506. The work of LR is supported by NSF grants PHY-1620806 and PHY-1915071, the Chau Foundation HS Chau postdoc award, the Kavli Foundation grant “Kavli Dream Team,” and the Moore Foundation Award 8342. The work of JL is supported by the Moore Foundation Award 8342 and Harvard University institutional funding.

%%%%%%%%%%%%%%%%%%%%%%%%%%%%%%%%%%
%%%%%%%%%%%%%%%%%%%%%%%%%%%%%%%%%%
\appendix
%%%%%%%%%%%%%%%%%%%%%%%%%%%%%%%%%%
%%%%%%%%%%%%%%%%%%%%%%%%%%%%%%%%%%

\section{Derivation of $\bar{\sigma} \sim m_{\sigma}^{3/2}$ and $\bar{\sigma} \sim m_{\sigma}^{9/4}$ scaling for $\Gamma_{\sigma} = m_{\sigma}^3/\Lambda^2$}\label{app:scaling}
%%%%%%%%%%%%%%%%%%%%%%%%
%%%%%%%%%%%%%%%%%%%%%%%%

In \Sec{sec:nsr_satisfied}, we saw that the $m_{\sigma}$-$\bar{\sigma}$ distribution that generates large $\fnl$ follows a clear scaling relation. We will derive the power of these scaling relations in this section.  We will first consider the case where the inflaton decays before curvaton oscillation, $\Gamma_{\phi}>m_{\sigma}$, meaning that the curvaton starts to oscillate during radiation domination. Right after the end of inflation, inflatons start to oscillate and redshift like matter. Curvatons, having a much smaller mass than the Hubble parameter at the end of inflation, are still stuck at the amplitude they had during inflation, and redshift like vacuum energy. Therefore, at the time of inflaton decay, which is defined by $H_{D\phi} \equiv \Gamma_{\phi}$, the inflaton energy density is
\beq
\begin{split}
\rho_{\phi, D\phi} &= \left(\frac{a_{\rm end}}{a_{D\phi}}\right)^3\rho_{\phi,{\rm end}}= \left(\frac{H_{D\phi}}{H_{\rm end}}\right)^{2}\rho_{\phi,{\rm end}}= \left(\frac{\Gamma_{\phi}}{H_{\rm end}}\right)^{2}\rho_{\phi,{\rm end}}= \Gamma_{\phi}^2 3M^2_{\text{pl}}.
\end{split}
\eeq
Meanwhile, the curvaton energy density is simply
\beq
\rho_{\sigma,D_{\phi}} = \rho_{\sigma, {\rm end}} = \frac{1}{2}m_{\sigma}^2 \bar{\sigma}^2.
\eeq

After inflaton decay, the universe becomes radiation dominated, and stays radiation dominated until at least curvaton oscillation at $H_{m_\sigma} = m_{\sigma}$. The energy density in the radiation fluid at the time of curvaton oscillation is
\beq
\begin{split}
\rho_{\gamma,m_{\sigma}} &= \left(\frac{a_{D\phi}}{a_{m_{\sigma}}}\right)^4\rho_{\phi, D\phi}= \left(\frac{H_{m_\sigma}}{H_{D\phi}}\right)^2 3M_{\text{pl}}^2\alpha_{\phi}^2m_{\phi}^2=m_\sigma^2 3M_{\text{pl}}^2.
\end{split}
\eeq 
Meanwhile, since the curvaton has not been oscillating,
\beq
\begin{split}
\rho_{\sigma, m_\sigma} = \rho_{\sigma, D_{\phi}} = \frac{1}{2}m_\sigma^2 \bar{\sigma}^2.
\end{split}
\eeq 

Once the curvaton begins to oscillate, it evolves as matter until it decays at $H = H_{D\sigma}=\Gamma_{\sigma}$. At the time of curvaton decay,
\beq
\rho_{\gamma,D_{\sigma}} = \left(\frac{a_{m_\sigma}}{a_{D_\sigma}}\right)^4 \rho_{\gamma, m_\sigma}
\eeq 
and 
\beq 
\rho_{\sigma,D_{\sigma}} = \left(\frac{a_{m_\sigma}}{a_{D_\sigma}}\right)^3 \rho_{\sigma, m_\sigma}. 
\eeq
Between curvaton oscillation and curvaton decay, the universe can be radiation dominated throughout, or transition to matter domination at some point. Since we know the energy density in radiation and curvaton at the time of curvaton decay, we can calculate when this matter-radiation equality happens,
\beq
\frac{a_{\rm eq}}{a_{m_{\sigma}}} = \frac{\rho_{\gamma,m_{\sigma}}}{\rho_{\sigma,m_{\sigma}}} = 6 \frac{\Mpl^2}{\bar{\sigma}^2}.
\eeq
Since, by definition, the universe is radiation dominated before matter-radiation equality, $H\propto a^{-2}$ during this period, and thus
\beq
\frac{H_{\rm eq}}{H_{m_{\sigma}}} = \frac{H_{\rm eq}}{m_{\sigma}}= \left(\frac{a_{\rm eq}}{a_{m_{\sigma}}}\right)^{-2} = \frac{\bar{\sigma}^{4}}{36\Mpl^4} \Rightarrow \frac{H_{\rm eq}}{\Gamma_{\sigma}} = \frac{\bar{\sigma}^{4}m_{\sigma}}{36\Mpl^4}\frac{\Lambda^2}{m_{\sigma}^3} = \frac{\bar{\sigma}^{4}\Lambda^2}{36\Mpl^4 m_{\sigma}^2}.
\eeq
For the range of $\bar{\sigma}$ and $m_{\sigma}$ relevant to our parameter scans in \Sec{sec:nsr_satisfied}, $H_{\rm eq}/\Gamma_{\sigma}$ is always less than 1, so the universe stays radiation dominated until curvaton decay. The radiation and curvaton energy densities at the time of curvaton decay are then
\beq
\begin{split}
\rho_{\gamma,D_{\sigma}} &= \left(\frac{a_{m_\sigma}}{a_{D_\sigma}}\right)^4 \rho_{\gamma, m_\sigma}= \left(\frac{H_{m_\sigma}}{H_{D_\sigma}}\right)^{-2} \rho_{\gamma, m_\sigma}= \alpha_{\sigma}^2 (3m_{\sigma}^2 \Mpl^2),
\end{split}
\eeq
and
\beq 
\begin{split}
\rho_{\sigma,D_{\sigma}} &= \left(\frac{a_{m_\sigma}}{a_{D_\sigma}}\right)^3 \rho_{\sigma, m_\sigma}= \left(\frac{H_{m_\sigma}}{H_{D_\sigma}}\right)^{-3/2} \rho_{\sigma, m_\sigma}= \alpha_{\sigma}^{3/2} \left(\frac{1}{2}m_{\sigma}^2 \bar{\sigma}^2\right),
\end{split}
\eeq
where $\alpha_{\sigma}\equiv \Gamma_{\sigma}/m_{\sigma}$. $\Omega_{\sigma_{D_\sigma}}$ is then
\beq 
\begin{split}
\Omega_{\sigma_{D_\sigma}} &\equiv \frac{\rho_{\sigma,D_{\sigma}}}{\rho_{\sigma,D_{\sigma}}+\rho_{\gamma,D_{\sigma}}}= \frac{1}{1+\frac{\rho_{\gamma,D_{\sigma}}}{\rho_{\sigma,D_{\sigma}}}} = \frac{1}{1+\frac{6 M_\text{pl}^2(\alpha_\sigma)^{1/2}}{\bar{\sigma}^2}}.
\end{split}
\eeq
and 
\begin{equation}
\begin{split}
R_{\sigma} &= \frac{1}{1+\frac{8 M_\text{pl}^2(\alpha_\sigma)^{1/2}}{\bar{\sigma}^2}}.
\end{split}
\end{equation}
In the limit where $R_{\sigma}\ll 1$, the non-Gaussianity is given by
\beq
\begin{split}
\fnl &\sim \frac{1}{R_{\sigma}}\Omega_{\sigma_{D\sigma}}^2\\
&\sim \frac{1}{R_{\sigma}}\left(\frac{R_{\sigma}^2 P_{S_{\sigma}}}{P_{\zeta_{\phi}}}\right)^2\\
&\sim \frac{\bar{\sigma}^2}{\alpha_{\sigma}^{3/2}},
\end{split}
\eeq
where on the last line we have only kept the dependence on $\bar{\sigma}$ and $m_{\sigma}$. For $\Gamma_{\sigma} = m_{\sigma}^3/\Lambda^2$, $\fnl \sim \bar{\sigma}^2/m_{\sigma}^3$, leading to the scaling relation $\bar{\sigma}\sim m_{\sigma}^{3/2}$, for any value of $\Lambda$.

Now we will consider the case where the curvaton oscillates before inflaton decay, $m_{\sigma}>\Gamma_{\phi}$. At the time of curvaton oscillation, the universe is dominated by the matter-like inflaton. The inflaton energy density is
\beq
\begin{split}
\rho_{\phi,\; m_{\sigma}} &= \left(\frac{a_{0}}{a_{m_{\sigma}}}\right)^3\rho_{\phi,{\rm end}}= \left(\frac{H_{m_{\sigma}}}{H_{\rm end}}\right)^{2}\rho_{\phi,{\rm end}}= \left(\frac{m_{\sigma}}{H_{\rm end}}\right)^{2}\rho_{\phi,{\rm end}}= 3m_{\sigma}^2M^2_{\text{pl}}, 
\end{split}
\eeq
and the curvaton energy density is,
\beq
\begin{split}
\rho_{\sigma, m_\sigma} = \frac{1}{2}m_\sigma^2 \bar{\sigma}^2.
\end{split}
\eeq
Right after inflaton decay ($H_{D\phi} = \Gamma_{\phi}$) occurs, we have
\beq
\begin{split}
\rho_{\gamma, D\phi} &= \left(\frac{a_{m_{\sigma}}}{a_{D_{\phi}}}\right)^3 \rho_{\phi,m_{\sigma}}= \left(\frac{H_{D_{\phi}}}{H_{m_{\sigma}}}\right)^2\left(3 m_{\sigma}^2 \Mpl^2 \right) = 3 \Mpl^2 \Gamma_{\phi}^2,
\end{split}
\eeq
and
\beq
\begin{split}
\rho_{\sigma,D_{\phi}} &= \left(\frac{a_{m_{\sigma}}}{a_{D_{\phi}}}\right)^3\frac{1}{2}m_\sigma^2 \bar{\sigma}^2= \left(\frac{H_{D_{\phi}}}{H_{m_{\sigma}}}\right)^2 \left(\frac{1}{2}m_\sigma^2 \bar{\sigma}^2\right)= \frac{1}{2}\bar{\sigma}^2 \Gamma_{\phi}^2.
\end{split}
\eeq

The universe is radiation dominated after the inflaton decay, and matter-radiation equality happens at
\beq
\frac{H_{\rm eq}}{H_{D\phi}} = \frac{H_{\rm eq}}{H_{D\phi}}= \left(\frac{a_{\rm eq}}{a_{D\phi}}\right)^{-2} =\left(\frac{\rho_{\gamma,D\phi}}{\rho_{\sigma,D\phi}}\right)^{-2} = \frac{\bar{\sigma}^{4}}{36\Mpl^4} \Rightarrow \frac{H_{\rm eq}}{\Gamma_\sigma} = \frac{\Gamma_{\phi}\bar{\sigma}^{4} \Lambda^2}{36\Mpl^4 m_\sigma^3}.
\eeq
For the range of $\bar{\sigma}$ and $\Gamma_{\phi}$ relevant to our parameter scan in \Sec{sec:nsr_satisfied}, $H_{\rm eq}/\Gamma_{\sigma}$ is always less than 1, so the universe stays radiation dominated until curvaton decay. The radiation and curvaton energy densities at the time of curvaton decay are then
\beq
\begin{split}
\rho_{\gamma,D_{\sigma}} &= \left(\frac{a_{D_{\phi}}}{a_{D_\sigma}}\right)^4 \rho_{\gamma, D_{\phi}}= \left(\frac{H_{D_{\phi}}}{H_{D_\sigma}}\right)^{-2}\left( 3 \Mpl^2 \Gamma_{\phi}^2\right) = \left(\frac{\Gamma_{\phi}}{\alpha_{\sigma}m_{\sigma}}\right)^{-2}\left( 3 \Mpl^2 \Gamma_{\phi}^2\right)= 3 \Mpl^2 (\alpha_{\sigma}m_{\sigma})^2,
\end{split}
\eeq and
\beq 
\begin{split}
\rho_{\sigma,D_{\sigma}} &= \left(\frac{a_{D_{\phi}}}{a_{D_\sigma}}\right)^3 \rho_{\sigma, D_{\phi}}= \left(\frac{H_{D_{\phi}}}{H_{D_\sigma}}\right)^{-3/2} \rho_{\sigma, D_{\phi}}= \left(\alpha_{\sigma}m_{\sigma}\right)^{3/2}(\Gamma_{\phi})^{1/2} \frac{1}{2}\bar{\sigma}^2.
\end{split}
\eeq
$\Omega_{\sigma_{D_\sigma}}$ is then
\beq 
\begin{split}
\Omega_{\sigma_{D_\sigma}} &= \frac{\rho_{\sigma,D_{\sigma}}}{\rho_{\sigma,D_{\sigma}}+\rho_{\gamma,D_{\sigma}}}= \frac{1}{1+\frac{\rho_{\gamma,D_{\sigma}}}{\rho_{\sigma,D_{\sigma}}}}= \frac{1}{1+\frac{6 M_\text{pl}^2}{\bar{\sigma}^2}\left(\frac{\alpha_{\sigma}m_{\sigma}}{\Gamma_{\phi}}\right)^{1/2}}.
\end{split}
\eeq
and 
\begin{equation}
\begin{split}
R_{\sigma} &= \frac{1}{1+\frac{8 M_\text{pl}^2}{\bar{\sigma}^2}\left(\frac{\alpha_{\sigma}m_{\sigma}}{\Gamma_{\phi}}\right)^{1/2}}.
\end{split}
\end{equation}
In the limit where $R_{\sigma}\ll 1$, the non-Gaussianity is given by
\beq
\begin{split}
\fnl &\sim \frac{1}{R_{\sigma}}\Omega_{\sigma_{D\sigma}}^2\\
&\sim \frac{1}{R_{\sigma}}\left(\frac{R_{\sigma}^2 P_{S_{\sigma}}}{P_{\zeta_{\phi}}}\right)^2\\
&\sim \frac{\bar{\sigma}^2}{\left(\alpha_{\sigma}m_{\sigma}\right)^{3/2}},
\end{split}
\eeq
where on the last line we have kept only the dependence on $\bar{\sigma}$ and $m_{\sigma}$. For $\Gamma_{\sigma} = m_{\sigma}^3/\Lambda^2$, $\fnl \sim \bar{\sigma}^2/m_{\sigma}^{9/2}$, leading to the scaling relation $\bar{\sigma}\sim m_{\sigma}^{9/4}$ for constant $\fnl$, for any value of $\Lambda$.

%%%%%%%%%%%%%%%%%%%%%
%%%%%%%%%%%%%%%%%%%%%
\bibliographystyle{utphys}
\bibliography{refs}
%%%%%%%%%%%%%%%%%%%%%
%%%%%%%%%%%%%%%%%%%%%

\end{document}